\documentclass{article}

\usepackage{arxiv}

\usepackage[utf8]{inputenc} % allow utf-8 input
\usepackage[T1]{fontenc}    % use 8-bit T1 fonts
\usepackage{hyperref}       % hyperlinks
\usepackage{url}            % simple URL typesetting
\usepackage{booktabs}       % professional-quality tables
\usepackage{amsfonts}       % blackboard math symbols
\usepackage{nicefrac}       % compact symbols for 1/2, etc.
\usepackage{microtype}      % microtypography
\usepackage{lipsum}
\usepackage{graphicx}

\title{Bankruptcy prediction using disclosure text features}

\author{
    Sridhar Ravula
   \\
    Department of Analytics \\
    Harrisburg University of Science and Technology \\
  Harrisburg, PA 17101 \\
  \texttt{\href{mailto:sravula@my.harrisburgu.edu}{\nolinkurl{sravula@my.harrisburgu.edu}}} \\
  }

\newlength{\csllabelwidth}
\newlength{\cslhangindent}
\newenvironment{cslreferences}%
  {\setlength{\parindent}{0pt}%
  \everypar{\setlength{\hangindent}{\cslhangindent}}\ignorespaces}%
  {\par}
 {% don't indent paragraphs
  \setlength{\parindent}{0pt}
  \ifodd #1 \everypar{\setlength{\hangindent}{\cslhangindent}}\ignorespaces\fi
  \ifnum #2 > 0
  \setlength{\parskip}{#3\baselineskip}
  \fi
 }%
 {}
\usepackage{calc} % for \widthof, \maxof

\begin{document}
\maketitle

\def\tightlist{}

\begin{abstract}
A public firm's bankruptcy prediction is an important financial research
problem because of the security price downside risks. Traditional
methods rely on accounting metrics that suffer from shortcomings like
window dressing and retrospective focus. While disclosure text-based
metrics overcome some of these issues, current methods excessively focus
on disclosure tone and sentiment. There is a requirement to relate
meaningful signals in the disclosure text to financial outcomes and
quantify the disclosure text data. This work proposes a new distress
dictionary based on the sentences used by managers in explaining
financial status. It demonstrates the significant differences in
linguistic features between bankrupt and non-bankrupt firms. Further,
using a large sample of 500 bankrupt firms, it builds predictive models
and compares the performance against two dictionaries used in financial
text analysis. This research shows that the proposed stress dictionary
captures unique information from disclosures and the predictive models
based on its features have the highest accuracy.
\end{abstract}


\hypertarget{introduction}{%
\section{Introduction}\label{introduction}}

Investors and analysts place great emphasis on security analysis and
valuation because of the potential excess returns on capital and the
downside risks. Research in this domain is potentially valuable because
market inefficiencies can result in volatility and crashes, costing the
economy billions of dollars. Analysts extensively use public firm's
disclosures as a source of information.

Investors are keen on knowing about the health of the firms they may
invest in the future. A firm in financial distress loses a significant
amount of its shareholder's value. If the management cannot tide over
the crisis, the firm may have to file for bankruptcy, resulting in a
50\% to 80\% loss of capital for shareholders and lenders. Financial
distress and bankruptcy prediction is an actively researched field.

Once a company is unable to come out of distress, it will become
insolvent. Insolvency is the state in which the company is not capable
of honoring some commitment. Lenders and claim holders can force the
insolvent company to discontinue operations. Managements file for
bankruptcy protection to recover from such a situation or liquidate it
in an orderly manner. Bankruptcy prediction has been an active research
topic for accounting researchers over decades. One of the pioneering
works Altman (1968) proposed the `Z score' model.

Investors and analysts traditionally depended on quantitative
information like accounting metrics for decision making. Multiple
attributes of these accounting metrics drove this trend. FACC and
accounting standards laid out what variables to be measured and
disclosed. Gathering, processing, and analyzing these quantitative
metrics was easy. Many free and commercial data providers automated the
data gathering and published these metrics. However, these metrics do
not always reveal the firm's current status and are not a good indicator
of the future. They suffer from shortcomings like window dressing and
retrospective focus.

Evidence exists for window dressing through commissions and omissions.
Rajan, Seru, and Vig (2015) showed that banks did not report information
regarding the deteriorating quality of borrowers' disclosures in the
run-up to the subprime crisis. Huizinga and Laeven (2012) said that
banks overstated the value of their distressed real estate assets and
regulatory capital. Window dressing, retrospective focus, and missing
variables impact models based on accounting metrics. Regulators and
investors who rely on such models have been impacted adversely in the
past due to model failures (Rajan, Seru, and Vig (2015)).

Another approach for bankruptcy prediction is using market-based
information. Classical efficient market theory and later option pricing
theories assume that all available information is reflected in market
prices. Under those conditions, accounting-based metrics do not have
additional information over and above market prices. More specifically,
a suitable market-based measure will reflect all available information
about bankruptcy probability. Hillegeist et al. (2004) developed a
prediction model based on market information, using option pricing
theory derived implied volatility. This model outperformed the Altman
(1968) z score model. Subsequently, numerous attempts have been made to
replicate these results. Wu, Gaunt, and Gray (2010) provides a
comparison of accounting and market-based models, along with others.
They conclude that the Hillegeist et al. (2004) model performs better
than the Z score model but is inferior to models that include
non-traditional metrics. Similarly, Tinoco and Wilson (2013) concluded
that accounting metrics based models and market-based models are
complimentary.

Hence researchers started paying more attention to alternative
approaches like textual analysis of disclosures. Management disclosures
have narrative content that contains important information. This
information can explain many firm attributes and organization outcomes,
and text analysis methods can extract this information. Prior works have
attempted to incorporate text features into accounting-based predictive
models. However, standalone text feature-based prediction models have
not been attempted. There is a need to understand how much information
can be extracted from disclosure texts and how useful, such information
is in predicting bankruptcy. This work addresses that gap.

\hypertarget{related-work}{%
\section{Related work}\label{related-work}}

Numerous researchers tried to explain various firm attributes using
disclosure narratives. Some analyzed MDA to explain future stock
performance (Tao, Deokar, and Deshmukh (2018)), future returns,
volatility, and firm profitability (Amel-Zadeh and Faasse (2016)),
bankruptcy (Yang, Dolar, and Mo (2018)), going-concern (Mayew,
Sethuraman, and Venkatachalam (2015),Enev (2017)), litigation risk
(Bourveau, Lou, and Wang (2018)), and incremental information over
earnings surprises, accruals and operating cash flows (OCF)(Feldman et
al. (2008),Feldman et al. (2010)).

Researchers attempted to incorporate text features into distress and
bankruptcy predictive models. Below is a brief review of the same.

Auditors express going-concern opinions based on the firm's obligations
and liquidity. Financial disclosures include these opinions. Change in
such disclosures can act as a signal to identify distress. However,
auditors do respond to external financial markets. Beams and Yan (2015)
examined the financial crisis's effect on auditor going-concern opinions
and concluded that the financial crisis led to increased auditor
conservatism. A going-concern opinion in disclosures is associated with
the number of forward-looking disclosures and their ambiguity. Enev
(2017) observed that while the absolute number of forward-looking
disclosures is lower for companies receiving a going concern opinion,
the proportion of forward-looking disclosures in the MDA is higher in
the presence of a going concern opinion. The results suggest generally
improved forward-looking disclosures in MDA when companies receive a
going concern opinion from their auditor.

One consequence of distress is financial constraints. Firms undergo
reduced cash flows during Stress, which results in liquidity events -
like dividend omissions or increases, equity recycling, and underfunded
pensions. Analysts measure the extent of financial constraints to assess
the capital structure. Bodnaruk, Loughran, and McDonald (2013) used a
constraining-words-based lexicon to measure the same. These measures
have a low correlation with traditional financial constraints measures
and predict subsequent liquidity events better. Ball, Hoberg, and
Maksimovic (2012) used text in firms' 10-Ks to measure investment delays
due to financial constraints. They found that the fundamental
limitations are the financing of R\&D expenditures rather than capital
expenditures and that the main challenge for firms is raising equity
capital to fund growth opportunities. These text-based measures predict
investment cuts following the financial crisis better than other indices
of financial constraints used in the literature.

Most prior bankruptcy prediction models were developed by using
financial ratios. However, signs of distress may appear in the
nonfinancial information earlier than changes in the financial ratios.
Current distress measures tend to miss extreme events, especially in the
banking sector (Gandhi, Loughran, and McDonald (2017)). In recent years,
qualitative information and text analysis have become necessary because
frequent changes in accounting standards have made it difficult to
compare financial numbers between years (Shirata et al. (2011)). Mayew,
Sethuraman, and Venkatachalam (2015) stressed the importance of
linguistic tone in assessing a firm's health. Using a sample of bankrupt
firms between 1995 and 2012, they concluded that management's opinion
about going-concern and the MDA's linguistic tone together predict
whether a firm will go bankrupt.

The language used by future bankrupt companies differs from non-bankrupt
companies. Hájek and Olej (2015) studied various word categories from
corporate annual reports and showed that the language used by bankrupt
companies shows stronger tenacity, accomplishment, familiarity, present
concern, exclusion, and denial. Bankrupt companies also use more modal,
positive, uncertain, and negative language. They built prediction models
combining both financial indicators and word categorizations as input
variables. This differential language usage is observed in non-English
firms' disclosures also. Shirata et al. (2011) analyzed the sentences in
Japanese financial reports to predict bankruptcy. Their research
revealed that the co-occurrence of words ``dividend'' or ``retained
earnings'' in a section distinguish between bankrupt companies and
non-bankrupt companies.

Working on U.S. Banks Gandhi, Loughran, and McDonald (2017) used
disclosure text sentiment as a proxy for bank distress. They found that
the annual report's more negative sentiment is associated with larger
delisting probabilities, lower odds of paying subsequent dividends,
higher subsequent loan loss provisions, and lower future return on
assets. Similarly, Lopatta, Gloger, and Jaeschke (2017) concluded that
firms at risk of bankruptcy use significantly more negative words in
their 10-K filings than comparable vital companies. This relationship
holds up until three years before the actual bankruptcy filing. Other
notable works using text analysis for bankruptcy prediction were Yang,
Dolar, and Mo (2018) and Mayew, Sethuraman, and Venkatachalam (2015).
Yang, Dolar, and Mo (2018) used high-frequency words from MDA and
compared the differences between bankrupt and non-bankrupt companies.
Mayew, Sethuraman, and Venkatachalam (2015) also analyzed MDA with a
focus on going-concern options. They found that both management's
opinion about ``going-concern'' reported in the MDA and the MDA's
linguistic tone together provide significant explanatory power in
predicting whether a firm will cease as a going concern. Also, the
predictive ability of disclosure is incremental to financial ratios,
market-based variables, even the auditor's going concern opinion and
extends to three years before the bankruptcy.

Most of the prior works focused on disclosure sentiment as an
incremental predictor for bankruptcy prediction. However, disclosure
text contains significantly more information other than sentiment, and
there is a need to extract and test its predictive power. To this end,
this quantitative correlation study evaluates the differences in
linguistic features between healthy and bankrupt disclosure texts.
Further, predictive models are built to assess the information content
and predictive power. The next section will outline the methods.

\hypertarget{method-and-materials}{%
\section{Method and materials}\label{method-and-materials}}

The prior sections have reviewed the literature and identified the gaps
in the text analysis of finance. As bankruptcy is a significant
organizational outcome for investors, this thesis focuses on the
bankruptcy prediction task. To this end, this quantitative correlation
study evaluates the differences in linguistic features between healthy
and bankrupt disclosure texts. Further, predictive models are built to
assess the information content and predictive power. This section
describes the framework, data analysis, and methodology used.

To summarize, this thesis has four key components.

\textbf{Text source: Management Discussion and Analysis from 10-K
disclosures.\\
Task: Bankruptcy prediction based on prior-year MDA.\\
Sample size: Balanced sample with 500 number of bankrupt and
non-bankrupt disclosures each.\\
Language models: Multiple, as described in later parts of this chapter.
}

The methods section consists of 4 sub-sections covering data, language
models, predictive models, and assessment criteria.

\hypertarget{sample-selection-criteria-and-data-sources}{%
\subsection{Sample selection criteria and data
sources}\label{sample-selection-criteria-and-data-sources}}

In this section, sample selection and data collection methods are
described. This work aims to extract knowledge from financial disclosure
text and use it for predictive tasks. It considers public listed
companies in the U.S. as the population. From 1994 to 2019, over 16,000
individual companies filed annual disclosures with SEC. New companies
get listed on exchanges through Initial Public Offering or corporate
spin-offs. Companies are delisted due to mergers, acquisitions, and
bankruptcies. As a result, there are \textasciitilde8000 listed public
companies in the year 2019 in the U.S.

This work focuses on bankruptcy prediction using disclosure text
characteristics. So, two samples are critical. One is a list of bankrupt
firms, and the other is a list of non-bankrupt firms.

\hypertarget{bankrupt-firms}{%
\subsubsection{Bankrupt firms}\label{bankrupt-firms}}

A critical component of this study is to identify firms that went
bankrupt. This work uses the list of bankrupt companies from the
UCLA-LoPucki Bankruptcy Research Database (BRD) maintained by LoPucki
(2006). UCLA School of Law collects, updates, and disseminates this
data. This dataset contains more than one-thousand large public
companies that have filed bankruptcy cases since October 1, 1979. BRD
defines a public company as a firm that filed an Annual Report (Form
10-K or form 10) with the SEC for a year ending not less than three
years before filing the bankruptcy case. BRD considers all firms with
more than \$ 100 million in assets in annual reports as ``large.''
Assets are measured in 1980 constant dollars (about \$ 3.1 current
dollar). Both Chapter 7 and Chapter 11 cases are included in the
bankruptcy list, whether filed by the debtors or creditors. From this
list, bankruptcies before 1994 are excluded. Since EDGAR maintains
online disclosures from 1994 onwards, it was convenient to extract those
filings. The exclusion of prior bankruptcies results in a new list of
\textasciitilde900 bankrupt companies. Around 7000 corresponding firm
filings exist in EDGAR. Companies without at least one prior year 10-K
filing are excluded from the list. Finally, the Management Discussion
and Analysis sections are extracted from these filings. A minimum
threshold of 100 words is used to filter out non-informative MDAs. This
filtering resulted in a sample of 500 company filings one year before
bankruptcy.

\hypertarget{non-bankrupt-firms}{%
\subsubsection{Non-Bankrupt firms}\label{non-bankrupt-firms}}

List of Non-Bankrupt firms is identified by starting with S\&P 1000 list
and excluding companies with bankruptcy history. The net result is 980
firms. Around 16000 filings exist for all these firms.

\hypertarget{sample-size}{%
\subsubsection{Sample size}\label{sample-size}}

Since annual bankruptcy incidence is less than 0.5\%, the number of all
filings one year prior to bankruptcies is very low, compared to
non-bankrupt filings. Hence, a balanced experiment design with an equal
number of bankrupt and non-bankrupt disclosures in the sample is used.
Five hundred non-bankrupt filings are randomly chosen from the
non-bankrupt filings.

\hypertarget{data-collection}{%
\subsection{Data collection}\label{data-collection}}

The method for the annual filings downloading has the below components.

\hypertarget{sec-data-extraction}{%
\subsubsection{SEC data extraction}\label{sec-data-extraction}}

From 1993 to 2018, Dec companies filed \textasciitilde20 million records
on EDGAR. For ease of access, SEC releases quarterly master indices for
the list of filings on EDGAR. This list has \textasciitilde{} 220,00
annual (10-K) filings relevant to this thesis. Custom R and python
scripts downloaded these 10-K documents programmatically.

\hypertarget{data-cleaning}{%
\subsubsection{Data cleaning}\label{data-cleaning}}

The text version of the filings on SEC is a collection of all files in a
submission. These include HTML, exhibits, jpg files, and XBRL files. A
fraction of the text file size will contain actual text. ASCII-encoded
pdfs, graphics, Xls, or other binary files can contribute to most of the
filings document size. The next processing step removed all non-text
content from disclosure documents, following Loughran and Mcdonald
(2009). These cleaned filings are stored in text format.

\hypertarget{mda-extraction}{%
\subsubsection{MDA extraction}\label{mda-extraction}}

For bankruptcy prediction, Management Analysis, and Discussion (MDA) is
the text features source. Management teams discussed the current firm
status and expected outcomes in the MDA section. A python script
extracted all the text between ``ITEM 7'' and ``ITEM 7 A''. Regular
expressions and combinations of these phrases are used to identify the
maximum number of the MDAs from 10-K files. In some disclosures, the MDA
section is ``incorporated by reference,'' referring to the shareholder's
annual report. The thesis included MDA material from the body of the
primary document. Also, it discarded all MDAs with less than a 100-word
count. Subsequent sections explain the generation of these texts'
numeric representation by using dictionary-based parsing or word
embeddings.

\hypertarget{variable-selection-and-language-models}{%
\subsection{Variable selection and language
models}\label{variable-selection-and-language-models}}

This subsection explains the dependent and independent variables used in
the thesis.

\hypertarget{outcome-definition}{%
\subsubsection{Outcome definition}\label{outcome-definition}}

The dependent variable for this thesis is a Bankruptcy filing. The
Bankruptcy filing Dummy equals one if the firm has filed for bankruptcy
protection within one year after the 10-K filing date, else 0.

\hypertarget{independent-variable-selection}{%
\subsubsection{Independent variable
selection}\label{independent-variable-selection}}

As outlined in the prior literature survey, numerous text representation
methods successfully extract information from financial disclosures.
However, often, they were used in combination with traditional
quantitative metrics and financial ratios. This thesis aims to identify
standalone information content in text and design methods for knowledge
extraction. This work evaluates numeric representations of MDA generated
using three types of Bag of Word dictionary-based language models. The
models are below.

\begin{enumerate}
\def\labelenumi{\arabic{enumi}.}
\tightlist
\item
  Linguistic Inquiry and Word Count (LIWC)
\item
  Loughran McDonalds Financial Dictionary (L.M.)
\item
  Stress Dictionary (S Dictionary)
\end{enumerate}

\hypertarget{dictionary-based-models.}{%
\subsubsection{Dictionary-based
models.}\label{dictionary-based-models.}}

Dictionary-based models are an extension of word frequency models. As
discussed in prior sections, word frequency models suffer from large
dimensionality and sparse matrix problems. One way to reduce the
dimensionality is to categorize words into different groups and compute
the category frequencies. These frequencies are normalized per thousand
words making comparison easier. These categorized word groups are called
dictionaries. Dictionary methods act as filters in extracting relevant
language features. For example, numerous words indicate negative
sentiment in a discourse. Collecting them under one group and computing
frequency helps in understanding document tone very quickly. These
advantages made dictionary-based methods prevalent in text analysis. The
next section covers the three dictionary-based models this thesis uses.

\hypertarget{liwc}{%
\paragraph{LIWC}\label{liwc}}

Linguistic Inquiry and Word Count (LIWC) is a text analysis program
developed by Pennebaker ( Pennebaker, Francis, and Booth (2001)). It
allows linguistic features analysis and content analysis. Also, the tool
can review stylistic aspects of language use across different contexts.
Since linguistic style reveals psychological information about a writer
and their underlying thinking, it is a useful tool in MDA analysis.
Researchers used LIWC in numerous financial text analysis studies.
Fisher, Garnsey, and Hughes (2016) provided a brief review.

LIWC examines written language and classifies it along up to 90 language
dimensions (Pennebaker et al. (2015)), including

1.Four summary language variables (analytical thinking, clout,
authenticity, and emotional tone)\\
2.Three linguistic descriptor categories (dictionary words, words per
sentence, six letters and above words)\\
3.Twenty-one standard language categories (e.g., articles, prepositions.
pronouns)\\
4.Forty-one psychological process word categories (e.g., affect,
cognition, biological processes, drives)\\
5.Six personal concern categories, five informal language markers, and
12 punctuation categories

The LIWC dimensions are hierarchically organized. For example, the word
`optimistic' falls into five categories: `optimism', `positive emotion',
`overall effect', `words longer than six letters' and `adjective'. The
program analyzes text files on a word-by-word basis, calculating the
number of words that match each of the 90 LIWC dimensions, expressed as
percentages of total words in the text, and records the data into one of
90 preset dictionary categories. The LIWC dictionary comprises over
6,000 words and stems. Each category is composed of a list of dictionary
words.

Several sources (e.g., Blogs, Expressive writing, Novels, Natural
Speech, NY Times, and Twitter) were used to form the dictionary. The
program classifies about 86 percent of the language used by people.
LIWC's external validity was tested; hence LIWC is a useful research
tool for measuring psychological processes, content analysis, and
assessing various linguistic features. LIWC measures for all MDAs are
generated using the LIWC2015 dictionary.

\hypertarget{lm-dictionary}{%
\paragraph{LM dictionary}\label{lm-dictionary}}

Loughran and Mcdonald (2011) demonstrated that word lists developed for
other disciplines misclassify common words in the financial text.
Loughran and Mcdonald (2011) created an alternative negative word list
(Fin-Neg) and five other word lists that better reflect tone in the
financial disclosures to overcome this. They tested the relation between
these word lists and 10 K filing returns, trading volume, return
volatility, fraud, material weakness, and unexpected earnings.
Subsequently, these word lists have been known as the L.M. dictionary,
and other researchers have used them in financial text analysis. Nguyen
and Huynh (2020), Gandhi, Loughran, and McDonald (2019). The five other
word lists are positive (Fin-Pos), uncertainty (Fin-Unc), litigious
(Fin-Lit), strong modal words (MW-Strong), and weak modal words
(MW-Weak). The Fin-Neg list has 2,337 words. This list includes
financial domain words that common negative words list exclude, i.e.,
restated, litigation, termination, discontinued, penalties, unpaid,
investigation, misstatement, misconduct, forfeiture, serious, allegedly,
noncompliance, deterioration, and felony. The Fin-Pos word list consists
of 353 words. The Fin-Unc list includes words indicating uncertainty and
has 285 words. For capturing propensity to litigate, 731 litigiousness
words are combined into the Fin-Lit list. It contains words such as
claimant, deposition, interlocutory, testimony, and tort. In the L.M.
dictionary, words from these three groups overlap. Strong and weak modal
words express levels of confidence. MW-Strong has 19 words: always,
highest, must, and will. MW-Weak has 27 words: could, depending, might,
and possibly.

For this work, Positive, Negative, and Uncertain words are included.
This work used the quanteda library, which includes the L.M. dictionary
(Benoit et al. (2018)), for generating numeric features.

\hypertarget{stress-dictionary}{%
\paragraph{Stress dictionary}\label{stress-dictionary}}

While LIWC and L.M. dictionaries extract the document's tone and
sentiment, they do not capture fundamental differences between bankrupt
and non-bankrupt companies. Also, L.M. demonstrated a need for task and
domain-specific dictionaries

Text features indicate differential language usage between bankrupt and
non-bankrupt companies. Distressed firms communicate the nature of
distress, remedial measures, and on-going concerns. Hence narrative of
distressed companies MDAs can differ from a healthy company MDA up to
three years before the bankruptcy For example, the following are some of
the statements from some distressed company's MDAs. \emph{``Operating
results are affected by indebtedness incurred to finance the acquisition
and by the amortization of capitalized fees and expenses incurred in
connection with such financing.''}\\
\emph{``The company is unlikely to be able to meet its cash flow needs
during..''}\\
\emph{``The company was downgraded in november 1994 by three primary
insurance rating agencies, and..''}\\
In a healthy firm's MDAs, we will not observe these sentences. The
following are some excerpts from healthy company MDAs.

\emph{``The increases in operating earnings were driven by revenue
growth and \ldots{}''}\\
\emph{``The company was in compliance with all debt covenants.''}\\
Further to the difference in content, the MDA content's linguistic
features in distressed firms can differ. This difference results from
obfuscation attempts - lengthy sentences describing the firm's state,
capturing the contingent conditions -narrating multiple agents'
attitudes, i.e., suppliers, lenders, economic factors, and management
prognosis.

The following statements highlight how a distressed firm communicates
its efforts in handling the situation \emph{``Since the company
currently does not have the means to repay the Series notes, management
is unable to predict the future liquidity of the company if the
restructuring is not accomplished.''}\\
\emph{``The company may be required to refinance such amounts as they
become due and payable. While the company believes that it will be able
to refinance such amounts, there can be no assurance that any Such
refinancing would be consummated or, if consummated, would be in An
amount sufficient to repay such obligations, particularly in light of
the company's high level of debt that will continue after the
Restructuring.''}\\
\emph{``After giving effect to this amendment, the company was in
compliance with the terms and restrictive covenants of its debt
obligations for fiscal 1994.''}\\
\emph{``The company has funded operations primarily from borrowings
under its debt agreements and the sale of its stock.''}\\
\emph{``The company was not in compliance with a net worth requirement
contained in its sale-leaseback agreement.''}\\
\emph{``As a result of the second quarter 1998 loss, the company was in
default of certain covenants based on ebitda.''}\\
\emph{``The loss incurred during the fourth quarter of the year ended
june 30, 1999 resulted in not being in compliance with the debt service
covenant''}\\
\emph{``The proposed plan currently contemplates the filing of a
pre-packaged chapter 11 plan of reorganization in order to\ldots{}''}\\
\emph{``These factors among others indicate that there is substantial
doubt about the company's ability to continue as a going concern.''}\\
\emph{``Considering our default of the loan agreements and our liquidity
as discussed above, there is substantial doubt about our ability to
continue as a going concern.''}

In contrast, healthy companies do not describe these details in a
lengthy manner. The following are excerpts from some healthy companies'
MDA \emph{``Management considers the company to be liquid and able to
meet its Obligations on both a short and long-term basis.'' }\\
\emph{``We had no amounts outstanding under our agreement.''}

The above observations suggest that a distress dictionary capturing
these differences would differentiate bankrupt and no-bankrupt firms.

\hypertarget{stress-dictionary-method.}{%
\paragraph{Stress dictionary method.}\label{stress-dictionary-method.}}

The dictionary is constructed using all MDAs from 2018. An MDA can
contain 5000 to 10,000 words. This study focuses on the ``liquidity and
capital requirements'' section, reported by most companies. The task is
to go through the words and identify the ones that may be red flags for
bankruptcy or Stress. The general decision criterion in the process is
high discriminatory power for identifying financial distress. The list
is prepared in two steps\\
1. Identification of differentiating words\\
2. Classifying the words into meaningful categories.\\
Similar to content analysis, which aims to extract information from the
text's tone, this work searches for words that might indicate debt
restructuring or distressed business situations. The first step
identified 70 candidate words. This list is refined in the next step.

\hypertarget{derivation-of-dictionary}{%
\paragraph{Derivation of dictionary}\label{derivation-of-dictionary}}

In a second step, we analyze the candidate list in detail. From the
preliminary list of 80 words, we categorize and select 70 words that are
consistent with prior literature.

\textbf{Category 1: Debt: Words used in expressing high indebtedness}

Companies deploy debt to take care of working capital and capital
expenditure requirements. During normal operations, firms manage debt
comfortably. When firms face difficulty in servicing the debt,
management discloses the status in MDA. This communication will result
in an increased frequency of debt-related words.\\
The following words characterize debt-related sentences: Agreement,
amendment, borrow, claim, collateral, guarantees, secured. A detailed
list is in appendix A.

\textbf{Category 2: Distress: Words used by companies close to
insolvency}

Companies in danger of bankruptcy exhibit several characteristics and
the MDA expresses the same. The expression of these characteristics
increases with an approaching need for bankruptcy filing. Debt covenant
violations are necessary pre-cursers to bankruptcy. Debt covenant
violations serve as early indicators to creditors, signaling potential
problems. Most violated covenants correspond to solvency (e.g., Interest
coverage and leverage), liquidity, and profitability requirements.
Managers try to avoid debt covenant default. Other words indicating
distress are loss, chapter 11, chapter 7, downgrade, and bankruptcy. We
add the following words to this list: covenant, default, breach,
violate, amend, restrictive, waiver

\textbf{Category 3: Restructure: Words used in restructuring sentences}

Managers try to manage distress through various mechanisms. Raising
fresh capital, debt restructuring, and selling of assets are some of
them. All these initiatives can be viewed as balance sheet restructuring
activities. MDA contains sentences explaining the proposed restructuring
activities. We add the following restructuring-related words to this
list: dispose, recapitalize, restructure, liquidate, alternative

\textbf{Category 4: Health: Characteristics of statements describing a
healthy state} Firms that are not at risk of bankruptcy express a
healthy state of the company in MDA. These sentences correspond to
solvency, profit, retained earnings, and dividend payment. We add the
following words to this list: retain, profit, cash, dividend, meet.
These four categories are defined as a dictionary and further used for
generating the numeric representation of MDAs.

\hypertarget{list-of-language-models}{%
\subsubsection{list of language models}\label{list-of-language-models}}

We built multiple combinations of language models from the different
language models described in the previous section. The final list of
language models is shown in table \ref{tab:langmodel}.

\begin{table}

\caption[Language models list]{\label{tab:langmodel}Language models used }
\centering
\begin{tabular}[t]{lll}
\toprule
Model & Name & Language Model\\
\midrule
Model 1 & LIWC & LIWC\\
Model 2 & LM & LM dictionary\\
Model 3 & Stress & Stress Dictionary\\
Model 4 & LIWC\_Stress & LIWC+ Stress Dictionary\\
Model 5 & LM\_Stress & LM Dictionary + Stress Dictionary\\
\addlinespace
Model 6 & LIWC\_LM\_Stress & LIWC+ LM dictionary + Stress Dictionary\\
\bottomrule
\end{tabular}
\end{table}

\hypertarget{predictive-modeling}{%
\subsection{Predictive modeling}\label{predictive-modeling}}

Once the documents have been transformed into numeric forms using
language models, they are fed into predictive models. The sample is
divided into two groups, bankrupt firms and non-bankrupt firms. The
outcome is a binary dependent variable. This binary outcome is modeled
using logistic regression, similar to a panel logit framework (Altman
and Hotchkiss (2010)).

\hypertarget{logit-model-specification}{%
\subsubsection{Logit model
specification}\label{logit-model-specification}}

Logistic regression is useful to model binary outcomes. It consists of a
logistic (logit) function and a binomial distribution. While standard
regression can be used to model binary outcomes, the model is not
interpretable. The outcome is not bounded, and an ad-hoc classification
rule is required to translate output to binary outcomes. Also, the
output cannot be converted to probabilities as, in some cases, the model
will produce estimates outside {[}0,1{]} bounds. The bounded constraint
can be overcome by modeling odds, i.e., \(p/(1-p)\). A log transform of
the odds will ensure that probabilities are symmetric at 0.5

The logistic function (also known as \emph{sigmoid function} or
\emph{inverse logit function}) critical ingredient of logistic
regression.

Logistic function:

\[f(x)=\frac{1}{1+e^{-x}}\] The logistic (logit) function:

\(1/(1+exp(-x))\)

Given log-odds: \(log(p/(1-p))\), logistic function is the inverse of
log-odds.

Another formula for logistic function:

\[g(x)=\frac{e^{x}}{e^{x}+1}\]

The logistic function, also called the sigmoid function, gives an `S'
shaped curve that can take any real-valued number (-\(\infty\) to
\(+\infty\)) and maps it to a value between 0 and 1.

\begin{figure}
\includegraphics[width=1\linewidth]{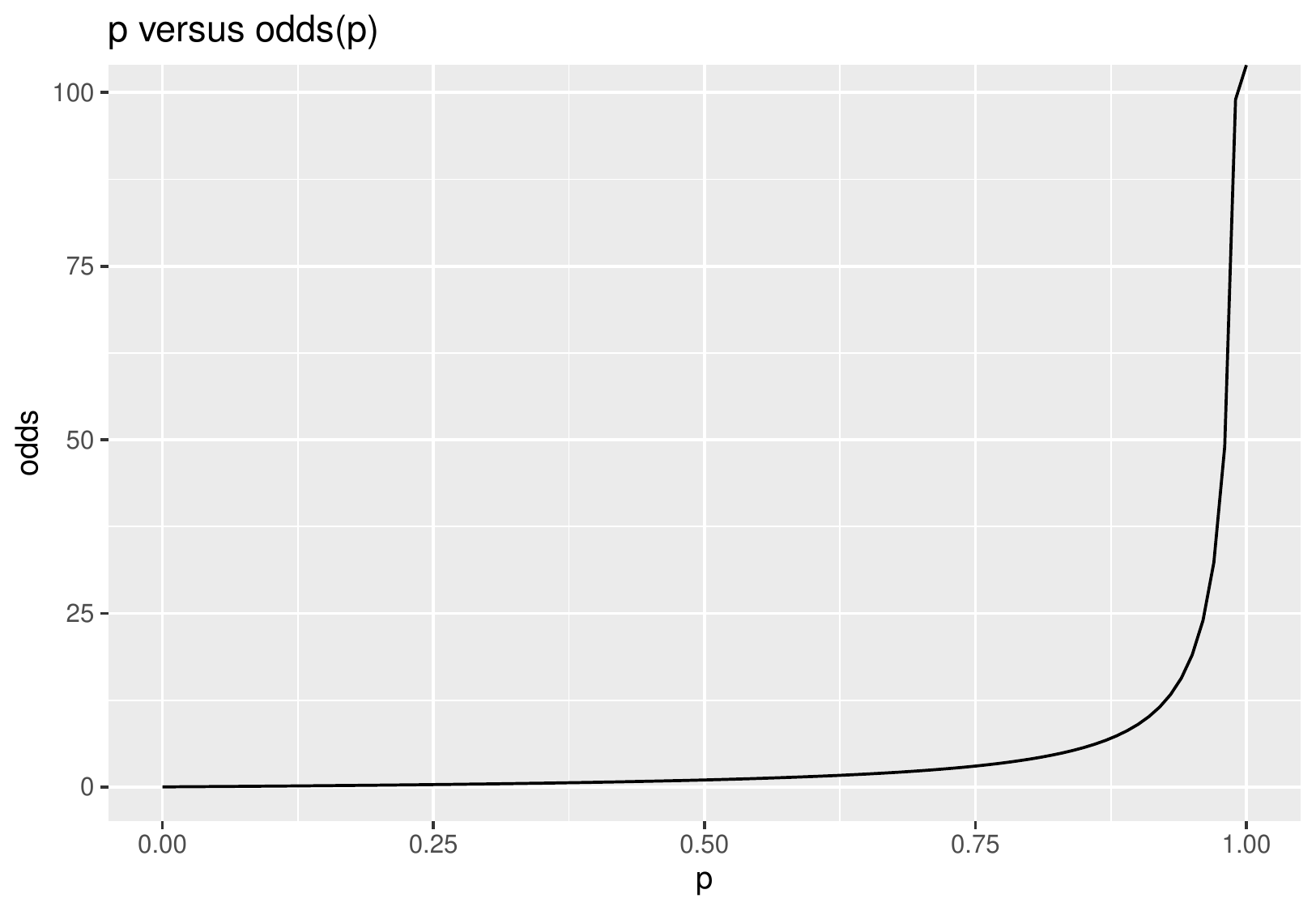} \caption{p versus odds(p)}\label{fig:OddsTransformation}
\end{figure}

\begin{figure}
\includegraphics[width=1\linewidth]{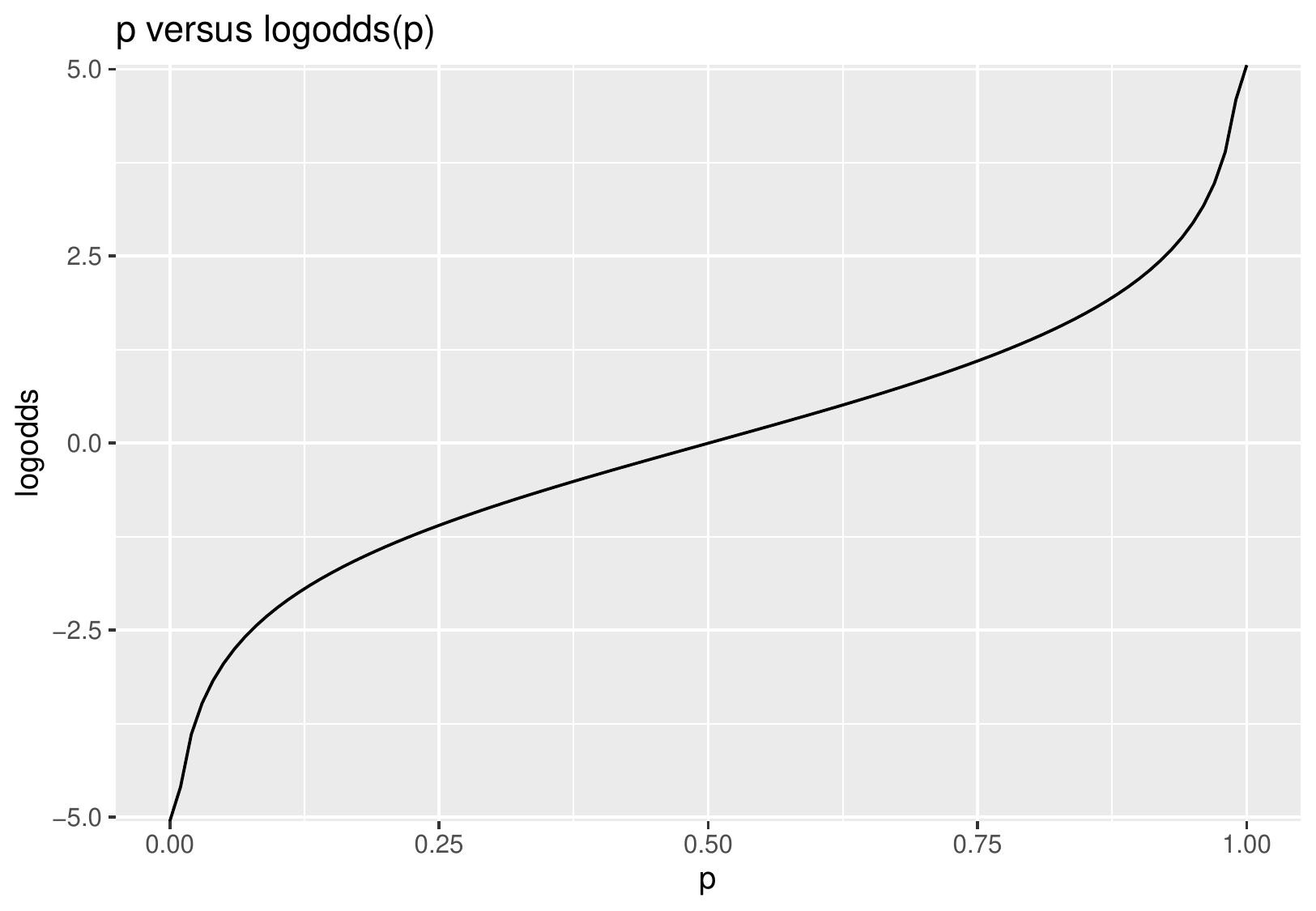} \caption{p versus logodds(p)}\label{fig:LogOddsTransformation}
\end{figure}

This transformation allows modeling a family of relationships between
continuous predictors and a binary outcome variable, in this case,
bankruptcy.

Key steps are

\begin{enumerate}
\def\labelenumi{\arabic{enumi}.}
\tightlist
\item
  Assuming that predictors are linearly related to the log-odds
\item
  Transform the odds to convert to probability
\item
  Estimate the data likelihood
\end{enumerate}

In this context, intercept shifts the curve left or right. Slopes make
the curve sharper or flatter, with respect to predictors. The logistic
starts at 0, ends at 1 and is symmetric around .5.

\begin{figure}
\includegraphics[width=1\linewidth]{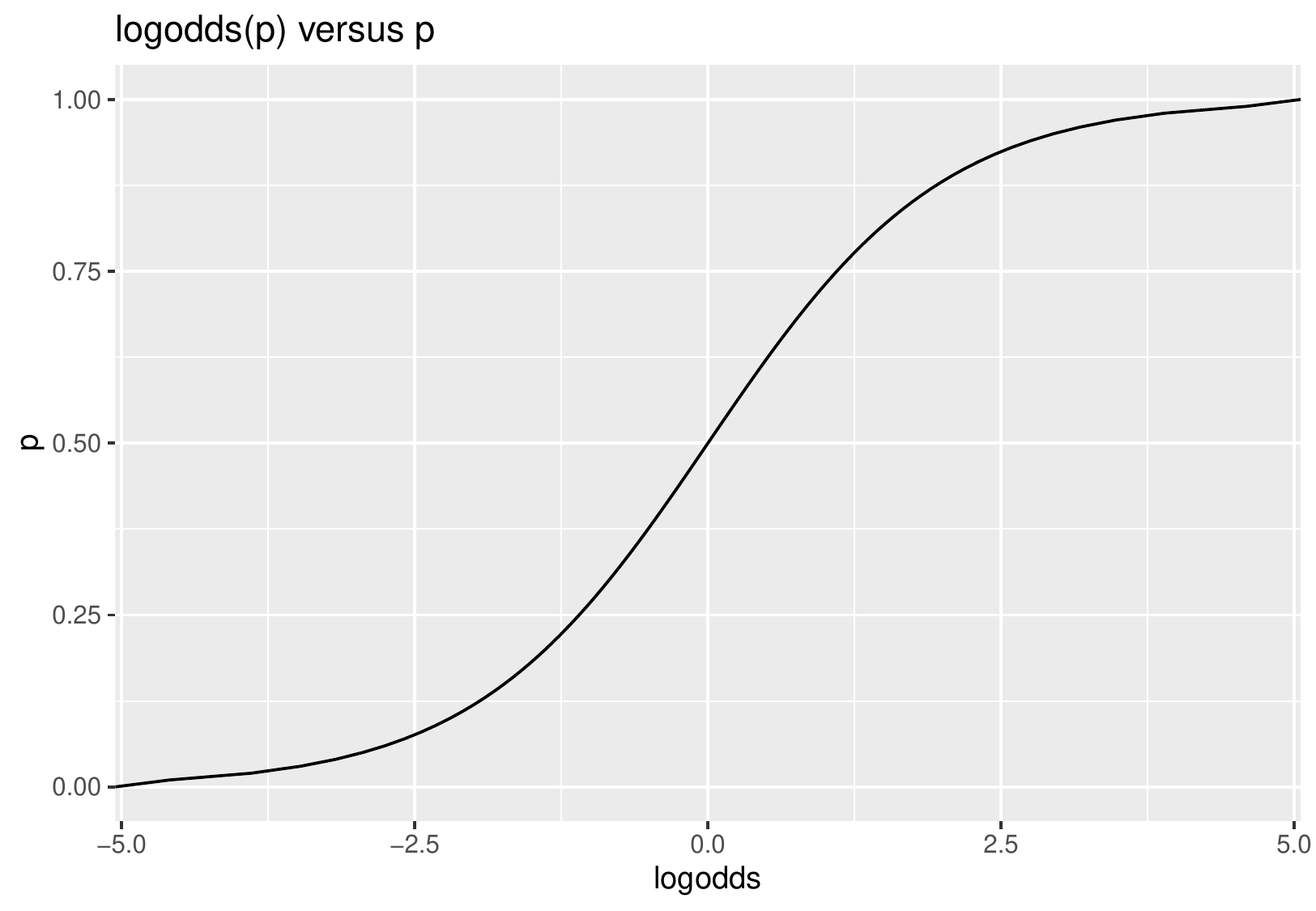} \caption{Logodds(p) vs p}\label{fig:Logistic}
\end{figure}

Logistic regression transforms the bankruptcy outcomes so that a linear
combination of predictors produces log-odds effects on the bankruptcy. A
model coefficient is transformed and interpreted as an odds multiplier.
These results are easily interpretable.

The logistic regression model used in this study is based on the
following mathematical definition. Bankruptcy variable coded using 1 and
0. \[
Y = 
\begin{cases} 
1&\text{bankrupt}\\
0&\text{non-bankrupt} 
\end{cases}
\]

Variable of interest

\[
p({\bf x}) = P[Y = 1 \mid {\bf X} = {\bf x}]
\]

\textbf{logistic regression} model.

\[
\log\left(\frac{p({\bf x})}{1 - p({\bf x})}\right) = \beta_0 + \beta_1 x_1 + \ldots  + \beta_{k - 1} x_{k - 1}
\]

This equation is similar to linear regression with \(k - 1\) predictors
for a total of \(k\) \(\beta\) parameters. Here, the left part of the
equation is the log odds. This gives the probability for a bankruptcy
\((Y = 1)\) divided by the probability of a non-bankruptcy \((Y = 0)\).
When the odds are 1, both events are equally likely. Odds greater than 1
indicate bankruptcy and vice versa.

\[
\frac{p({\bf x})}{1 - p({\bf x})} = \frac{P[Y = 1 \mid {\bf X} = {\bf x}]}{P[Y = 0 \mid {\bf X} = {\bf x}]}
\]

\hypertarget{evaluating-predictive-models}{%
\subsection{Evaluating predictive
models}\label{evaluating-predictive-models}}

Researchers evaluate Bankruptcy prediction models using multiple
criteria. In this thesis, we use Accuracy tables, the receiver operating
characteristics (ROC) curves, and information content tests. While ROC
curves inform forecasting accuracy, sensitivity, and specificity,
Information content tests evaluate the bankruptcy-related information
carried by the distress risk measures. The following section presents
the method of each.

\hypertarget{classification-accuracy-tables}{%
\subsubsection{Classification accuracy
tables}\label{classification-accuracy-tables}}

A perfect model classifies all observations accurately. Real models make
mistakes in classification. One way to evaluate the model's performance
is its misclassification rate. Alternatively, models accuracy can be
used, which measures the proportion of correction classifications

\[
\text{Misclassification }(\hat{C}, \text{Data}) = \frac{1} {n}\sum_{i = 1}^{n}I(y_i \neq \hat{C}({\bf x_i}))
\]

\[
I(y_i \neq \hat{C}({\bf x_i})) = 
\begin{cases} 
  0 & y_i = \hat{C}({\bf x_i}) \\
  1 & y_i \neq \hat{C}({\bf x_i}) \\
\end{cases}
\]

This measure is not useful in training data. This metric improves with
the number of parameters and hence will be biased towards large models.
This bias encourages overfitting. Hence, this metric needs to be
computed on test data, unseen by the model during training.

Accuracy tables can be further split into confusion matrix, to
understand the nature of misclassification. Confusion matrix categorizes
the classification errors into false negatives and false positives.

\begin{figure}
\includegraphics[width=1\linewidth]{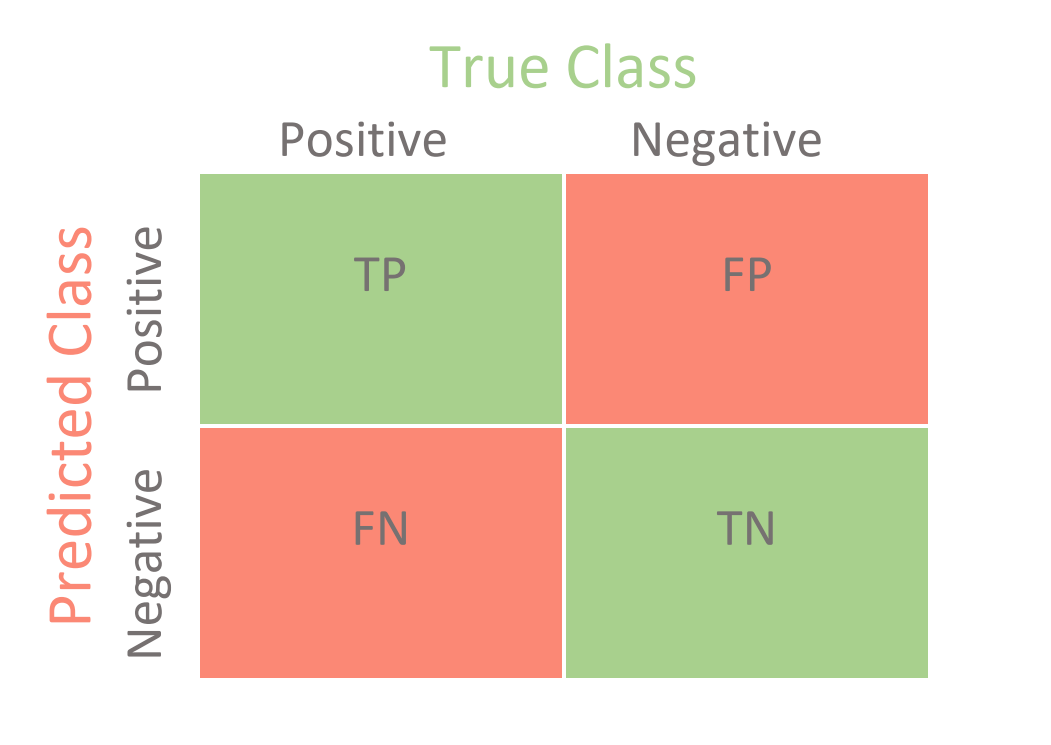} \caption{Confusion Matrix}\label{fig:ConfusionMatrix}
\end{figure}

Setting the classification threshold as 0.5 \[
\eta({\bf x}) = 0 \iff p({\bf x}) = 0.5
\]

Predictions can be used to create a confusion matrix as below.

\[
\text{Prev} = \frac{\text{P}}{\text{Total Obs}}= \frac{\text{TP + FN}}{\text{Total Obs}}
\]

A reasonable classifier has to outperform a naïve classifier that labels
all observations as majority class. In this work, a model classifying
every company as non-bankrupt will be the baseline. Apart from accuracy,
specificity and sensitivity can be used to evaluate models. Sensitivity
is the true-positive rate. Higher sensitivity means the model classifies
more positives correctly, reducing the false negatives. Specificity is
the true negative rate. Higher specificity means the classifier is
labeling true negatives correctly, reducing false positives. The
formulae are given below. \[
\text{Sensitivity} = \text{True Positive Rate} = \frac{\text{TP}}{\text{P}} = \frac{\text{TP}}{\text{TP + FN}}
\]

\[
\text{Specificity} = \text{True Negative Rate} = \frac{\text{TN}}{\text{N}} = \frac{\text{TN}}{\text{TN + FP}}
\]

Both metrics can be computed directly from the confusion matrix.

\textbf{Relationship between Accuracy, Specificity, and Sensitivity}

As we compute specificity and sensitivity from the confusion matrix,
different classification thresholds generate multiple sensitivity/
specificity values. It is normal to use 0.5 probability as ``cutoff.''
By modifying the cutoff, we can improve the sensitivity or specificity
at the overall accuracy expense. Also, if sensitivity improves,
specificity deteriorates, and vice versa.

\[
\hat{C}(\bf x) = 
\begin{cases} 
      1 & \hat{p}({\bf x}) > c \\
      0 & \hat{p}({\bf x}) \leq c 
\end{cases}
\]

\hypertarget{test-of-predictive-ability-receiver-operating-characteristics-roc}{%
\subsubsection{Test of predictive ability: Receiver Operating
Characteristics
(ROC)}\label{test-of-predictive-ability-receiver-operating-characteristics-roc}}

Receiver Operating Characteristics (ROC) curve is a method to assess the
accuracy of a continuous measurement for predicting a binary outcome. It
is used extensively in the Lifesciences Domain. Over the past two
decades, it gained acceptance as a bankruptcy prediction model
validation tool (Sobehart and Keenan (2001)).

For a bankruptcy prediction model, for a fixed cutoff c, we can compute
accuracy metrics and two types of classification errors: false negatives
and false positives. In bankruptcy prediction, the model generates the
measure of firm distress M, based on independent variables. This measure
is a continuous measurement. We derive a 1 (test positive)
classification as M exceeding a fixed threshold c: M\textgreater c.~For
bankruptcy detection, the binary outcome B, a good outcome of the test,
is when classification is 1 (the test is positive) among bankrupt
companies B=1. A bad outcome is when the classification is 1 (test is
positive) among non-bankrupt companies B=0. The true-positive fraction
is the probability of an estimated positive among the bankrupt firms:
TPF(c)=P\{M\textgreater c\textbar B=1\}. This value is the sensitivity
at cutoff c. Similarly, the false-positive fraction is the probability
of a bankrupt classification among the non-bankrupt firms: FPF(c)=
P\{M\textgreater c\textbar B=0\} ROC curve is the plot of TPF against at
various cutoff levels c.~It has FPF(c) on the x-axis and TPF(c) along
the y-axis.

A perfect bankruptcy prediction model - that is, the ranking on default
probability at cutoff c is equal to the ranking of failures at c --
would be able to capture all bankruptcies. This model corresponds to a
vertical line at 0 FPF. A random bankruptcy prediction model -- that is,
the ranking at cutoff c is not correlated with the ranking of failures
-- would have the same percentage of failures across each cutoff level.
This model corresponds to a line at 45\textsuperscript{0} to the x-axis.
Since we expect a bankruptcy prediction model to be better than a random
model, the ROC curve is expected to be between the perfect and the
random model.

To compare the two models' predictive ability, we calculate the area
under the ROC curve (AUC). Sobehart and Keenan (2001) used the AUC is
the decisive indicator for default model accuracy.

\hypertarget{information-content-test}{%
\subsubsection{Information Content
Test}\label{information-content-test}}

Information content tests help examine the proposed bankruptcy
prediction models. They evaluate if bankruptcy prediction models carry
more information than another set of variables. The use of Information
content tests has many precedents in bankruptcy prediction. They
complement the ROC curve analysis since (i) ROC curve analysis provides
users with a binary option, but users may not be making such decisions.
Users of bankruptcy prediction models are interested in determining
credit terms or portfolio weights. (ii) ROC curve analysis ignores
associated error costs based on context-specific type I/ type II errors.

There are two primary information criteria: the Akaike information
criterion (AIC) and the Bayes information criterion (BIC). When models
are built using the same data by maximum likelihood, smaller AIC or BIC
indicates a better fit.

\hypertarget{akaike-information-criterion}{%
\paragraph{Akaike Information
Criterion}\label{akaike-information-criterion}}

The AIC is the simpler of the two; it is defined as AIC = -2LL + 2k, in
which -2LL is the deviance (described below), and k is the number of
predictors in the model.

The maximum log-likelihood of a regression model is:
\([ \log L(\boldsymbol{\hat{\beta}}, \hat{\sigma}^2) = -\frac{n}{2}\log(2\pi) - \frac{n}{2}\log\left(\frac{\text{RSS}}{n}\right) - \frac{n}{2}, ]\)

Where \(\boldsymbol{\hat{\beta}}\) and \(\hat{\sigma}^2\) and
\(\text{RSS} = \sum_{i=1}^n (y_i - \hat{y}_i) ^ 2\) were selected to
maximize the likelihood.

From the above, AIC is derived as the difference between penalty and
log-likelihood

\([ \text{AIC} = -2 \log L(\boldsymbol{\hat{\beta}}, \hat{\sigma}^2) + 2k = 2k + n + n \log(2\pi) + n \log\left(\frac{\text{RSS}}{n}\right), ]\)

AIC combines two components of the model, i.e., the likelihood -- a
measure of ``goodness-of-fit'' and the penalty- proportional to the
model size. The likelihood portion of AIC for two models fit on the same
dataset is a function of RSS. Higher RSS (squared deviation) indicates a
poor model fit. A good model has low RSS and AIC. The penalty component
of AIC is \([ 2k, ]\), a function of the number of \(\beta\) parameters
used in the model. As k increases, AIC increases. A good model with a
small AIC will have a balance between the goodness of fit and uses a
small number of parameters.

\hypertarget{BIC}{%
\paragraph{Bayesian Information Criterion}\label{BIC}}

The BIC is similar to AIC but adjusts the penalty included by the number
of cases: BIC = -2LL + k x log(n) in which n is the number of cases in
the model. This way, BIC picks smaller models for larger sample sizes,
compared to AIC. For model selection, we use the model with the smallest
BIC.

\([ \text{BIC} = \log(n) k -2 \log L(\boldsymbol{\hat{\beta}}, \hat{\sigma}^2) = \log(n)k. + n + n\log(2\pi) + n\ log\left(\frac {\text{RSS}}{n}\right)]\)

The penalty for AIC is 2k whereas for BIC, it is \([ \log(n) p. ]\). For
datasets with \(log(n) > 2\) , BIC penalty will be higher compared to
AIC. Hence BIC will prefer smaller models for similar log-likelihoods.

\hypertarget{results}{%
\section{Results}\label{results}}

This research work focuses on building bankruptcy prediction models
using financial disclosures text features. Statistical analysis has been
conducted, and models are built as per the methodology described in
section \ref{method-and-materials}. This chapter will describe the
results.

The chapter is structured into multiple sub-sections covering
descriptive statistics of linguistic features, relationship with
bankruptcy, model performance, and evaluation.

\hypertarget{descriptive-statistics}{%
\subsection{Descriptive statistics}\label{descriptive-statistics}}

This section describes the statistical properties of datasets and
features used in this work

\hypertarget{bankrupt-companies}{%
\subsubsection{Bankrupt companies}\label{bankrupt-companies}}

The list of bankruptcies from LoPucki (2006) has more than 1000
observations. This dataset covers large bankruptcies from 1980 to date.

\begin{figure}
\includegraphics[width=1\linewidth]{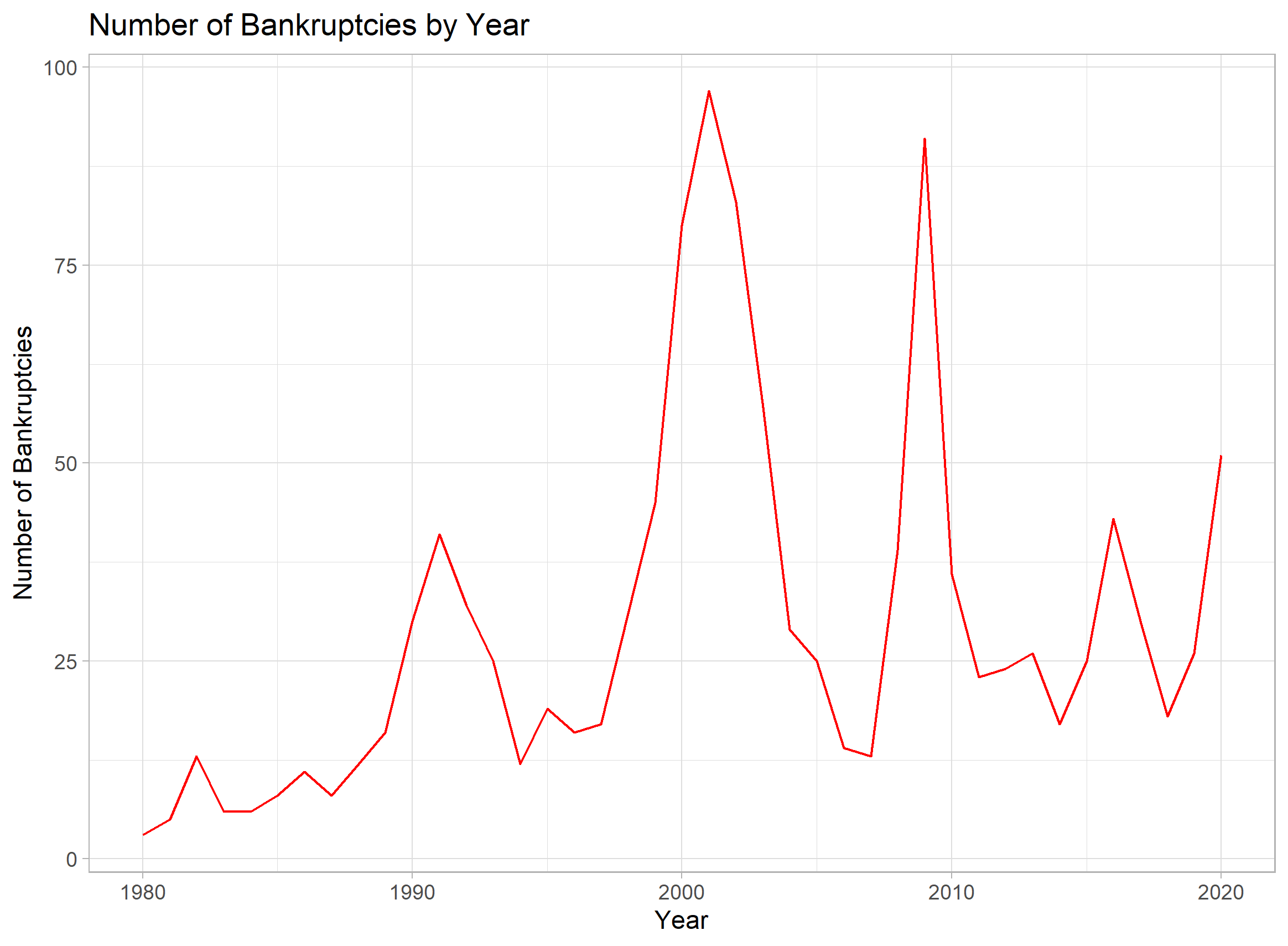} \caption{Number of bankruptcies filed by year}\label{fig:AnnualBankruptciesPlot}
\end{figure}

The annual bankruptcy filings trend is given in figure
\ref{fig:AnnualBankruptciesPlot}.

--\textgreater{}

On average, 29 firms filed for bankruptcy in a year, with median annual
bankruptcies at 25. A maximum of 97 bankruptcies was filed in the year
2001. Recall that this research has included bankruptcies till 2018
Dec.~

\hypertarget{mda-linguistic-features}{%
\subsubsection{MDA linguistic features}\label{mda-linguistic-features}}

For the selected bankrupt firms and healthy firms, all available MDAs
are transformed into numeric forms using three dictionaries, i.e., LIWC,
L.M., and stress dictionary. These linguistic features are averaged at
the group level and presented in the below table.

\begin{table}

\caption[Linguistic features summary]{\label{tab:LinguisticFeatures}Linguistic features words percentage }
\centering
\begin{tabular}[t]{lrrr}
\toprule
Feature & All & Bankrupt & Healthy\\
\midrule
WPS & 26.91 & 27.35 & 26.72\\
WC & 10109.29 & 10164.10 & 10085.28\\
Sixltr & 27.93 & 27.87 & 27.96\\
Dic & 162.26 & 160.67 & 162.96\\
function. & 30.57 & 30.67 & 30.52\\
\addlinespace
affect & 2.83 & 2.83 & 2.84\\
social & 3.37 & 3.29 & 3.41\\
cogproc & 7.38 & 7.19 & 7.47\\
percept & 0.30 & 0.32 & 0.29\\
bio & 0.98 & 0.97 & 0.98\\
\addlinespace
drives & 6.62 & 6.41 & 6.70\\
relativ & 10.71 & 10.70 & 10.72\\
AllPunc & 12.35 & 12.77 & 12.17\\
focuspast & 1.73 & 1.78 & 1.70\\
focuspresent & 2.69 & 2.61 & 2.73\\
\addlinespace
focusfuture & 0.78 & 0.80 & 0.78\\
anger & 0.04 & 0.03 & 0.04\\
posemo & 2.10 & 2.12 & 2.09\\
negemo & 0.73 & 0.71 & 0.74\\
debt & 2.72 & 3.04 & 2.58\\
\addlinespace
distress & 0.24 & 0.33 & 0.21\\
restructure & 0.08 & 0.09 & 0.07\\
healthy & 0.54 & 0.55 & 0.54\\
negative & 1.01 & 1.07 & 0.98\\
positive & 0.51 & 0.48 & 0.52\\
\addlinespace
uncertainty & 0.96 & 0.91 & 0.98\\
\bottomrule
\end{tabular}
\end{table}

Column ``All'' documents the summary statistics for all sample firms.
WPS and W.C. indicate that the sample firms' MDAs are in general lengthy
with \textasciitilde10000 words and 27 words per sentence, indicating
\textasciitilde400 sentences per MDA. Excluding WPS and W.C., all other
values are in percentages. Close to 30\% of words are complex (27.93
Sixltr ) and functional (function. 30.57). The sample firms' MDAs are
present-focused (focuspresent 2.69), and their future focus is one-third
of the present focus. Per LIWC classification, on average, the MDAs have
three times more positive words compared to negative words (posemo:2.1,
negemo: 0.73). Cognitive process-related and drives related words are
observed with similar frequency (cogproc:7.38, drives:6.62), while
Social/ affect words occur at half of that (social:3.37, affect:2.83).
Based on the L.M. dictionary, we can observe that negative and uncertain
words are double that of positive words frequency (negative:1.01,
uncertain:0.96, positive: 0.51) Stress dictionary features indicate that
debt words are prevalent at 2.72. In a typical MDA of 10,000 words
length, this indicates 270 words describing debt-related discussion and
disclosures. Distress and restructure related occur less frequently,
which can be expected as they are infrequent organizational outcomes.

The table's focus is Column ``Bankrupt,'' as it illustrates the summary
statistics of bankrupt firms. Bankrupt firms are more past focused. They
also use less cognitive and drives related words. A striking difference
is observed in the increase in debt and distress related word frequency.
They also show increased negative word frequency.

\begin{landscape}

\begin{figure}
\includegraphics[width=1\linewidth]{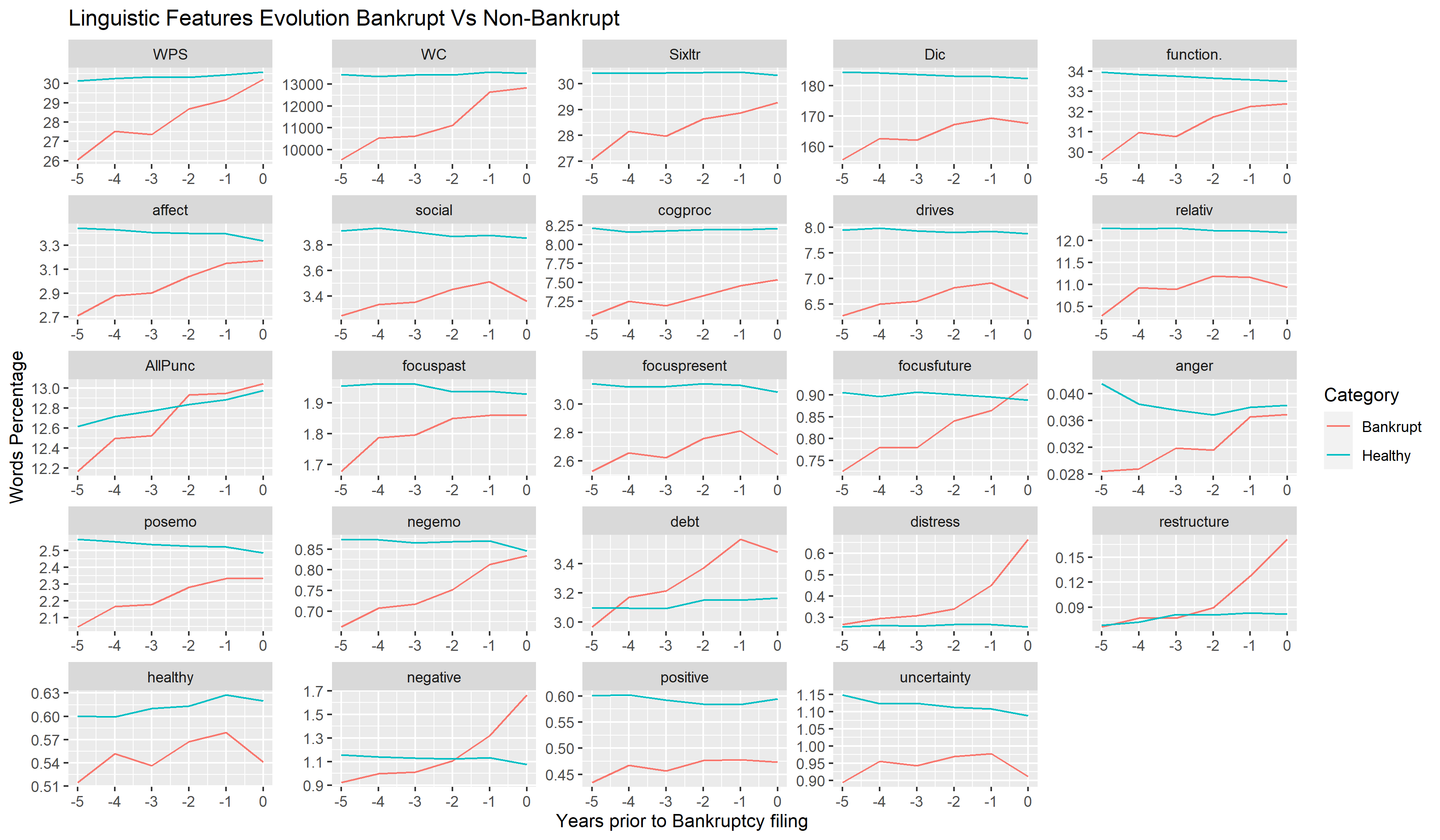} \caption{Linguistic features evolution}\label{fig:LinguisticFeaturesEvolution}
\end{figure}

\end{landscape}

Since we are interested in building predictive models using prior year
filings, it would be critical to observe how the linguistic features
trend for bankrupt companies compared to non-bankrupt companies. The
figure \ref{fig:LinguisticFeaturesEvolution} shows the same.

This figure depicts various types of word frequencies for bankrupt
companies during the year of bankruptcy and five prior years. For
comparison, sample non-bankrupt firms' word percentages are plotted over
six years, going back from the latest filing. The values are averaged
for bankrupt and non-bankrupt firms.

\textbf{Notable trends in LIWC features}\\
All linguistic LIWC features for bankrupt firms are lower than
non-bankrupt firms throughout the period. There is a gradual increase in
focuspast and focusfuture.

\textbf{Notable trends in L.M. features}\\
Bankrupt companies have lower uncertain and positive words throughout
the period. Negative words stat increasing two years before the
bankruptcy.

\textbf{Notable trends in Stress Dictionary features}

Stress dictionary features captured the evolution of distress and
bankruptcy. Debt related words exceed relative to healthy firms four
years before bankruptcy and gradually inch up further till the event of
bankruptcy. Distress related words remain marginally higher from 5 years
to 2 years before bankruptcy and dramatically increase after that.
Restructure related word frequency for bankrupt firms is
indistinguishable till two years before the bankruptcy. This observation
is expected as firms do not take up such costly exercises unless the
financial distress is unmanageable and covenant default is imminent.
There is no change in ``healthy'' frequency for both bankrupt and
non-bankrupt firms, though bankrupt firms have lower occurrence
throughout the period.

Overall, we can observe sufficient differences between bankrupt and
non-bankrupt firms.

\hypertarget{correlation-between-linguistic-features}{%
\subsubsection{Correlation between linguistic
features}\label{correlation-between-linguistic-features}}

Figures \ref{fig:LIWCCorrelations},
\ref{fig:LinguisticCorrelationsLMStress},\ref{fig:LinguisticCorrelations}
show correlation structure among LIWC features, LM-Stress dictionary and
selected variables from these three models.

Of the LIWC features, few are highly correlated, i.e., dictionary,
functional, social, and drives. All other features have low correlations
indicating they are capturing different information. In the stress
dictionary, debt and distress show a 0.45 correlation, which is
expected. Other variables are uncorrelated. Also, there is no
correlation between the L.M. dictionary and stress dictionary features.
Finally, selected variables from these three models are checked for
correlation. There is an insignificant correlation indicating minimum
overlap. This low correlation indicates their complementary nature, and
a hybrid model combining these features might perform better than
standalone models.

\begin{landscape}

\begin{figure}
\includegraphics[width=1\linewidth]{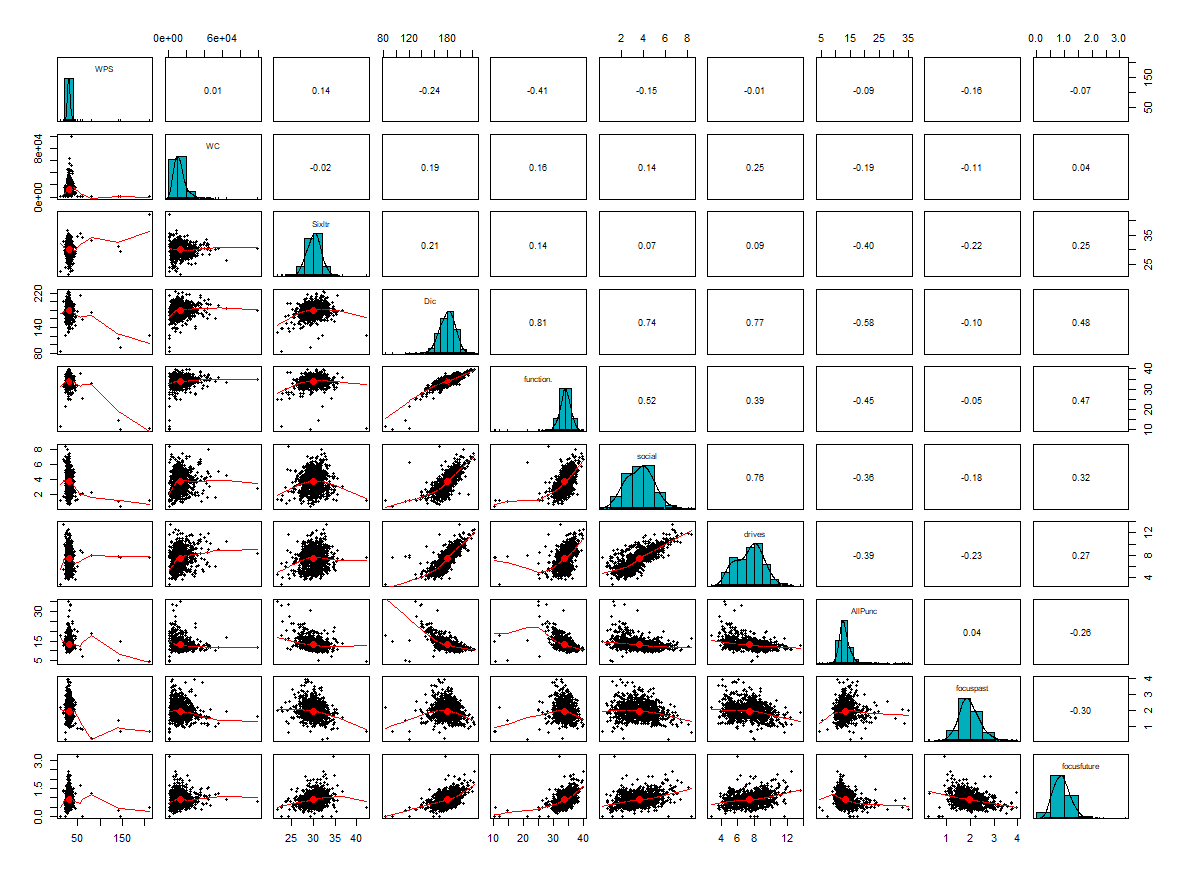} \caption{Correlations between LIWC features}\label{fig:LIWCCorrelations}
\end{figure}

\begin{figure}
\includegraphics[width=1\linewidth]{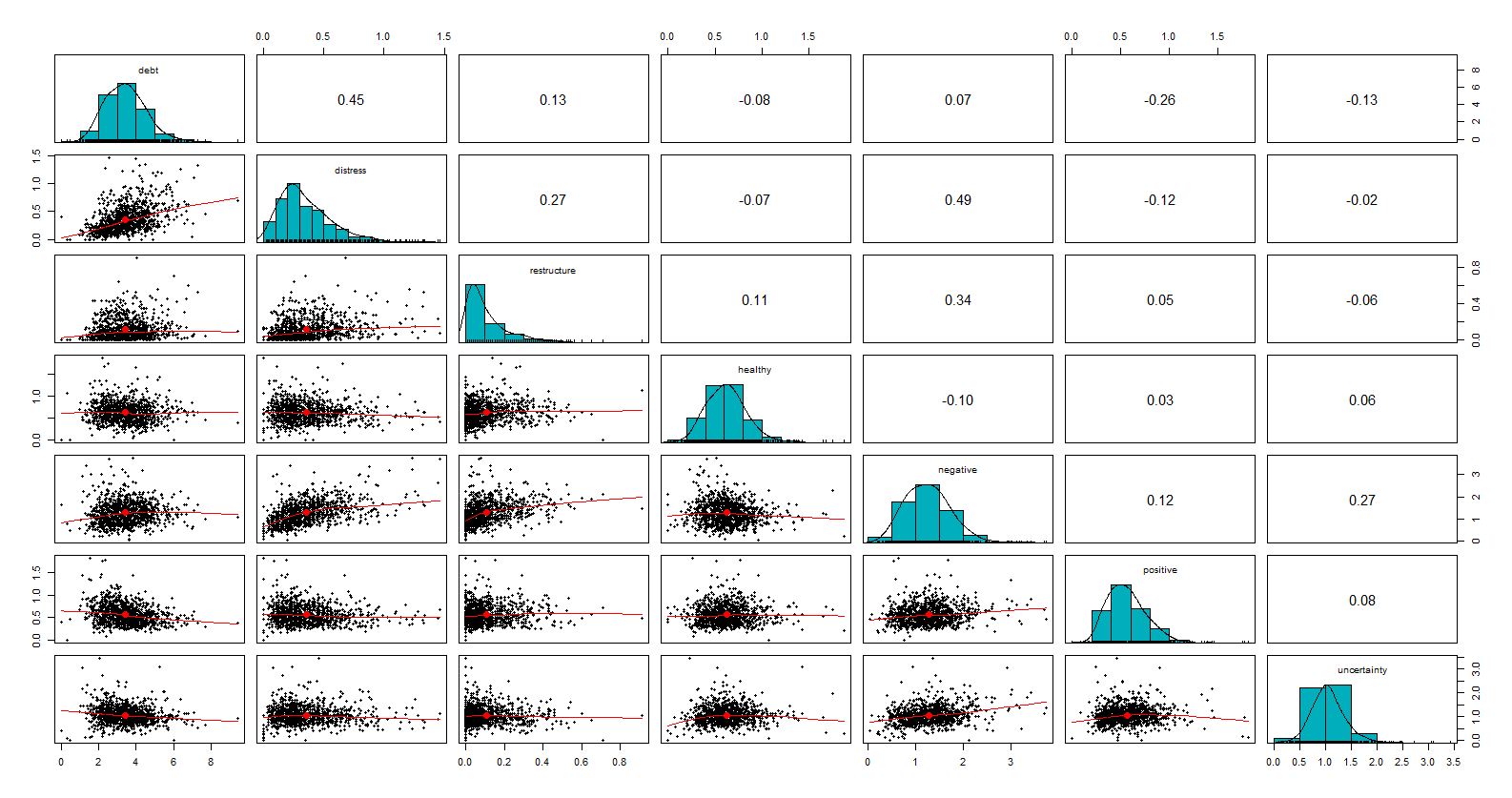} \caption{Correlations between LM and stress dictionary}\label{fig:LinguisticCorrelationsLMStress}
\end{figure}

\begin{figure}
\includegraphics[width=1\linewidth]{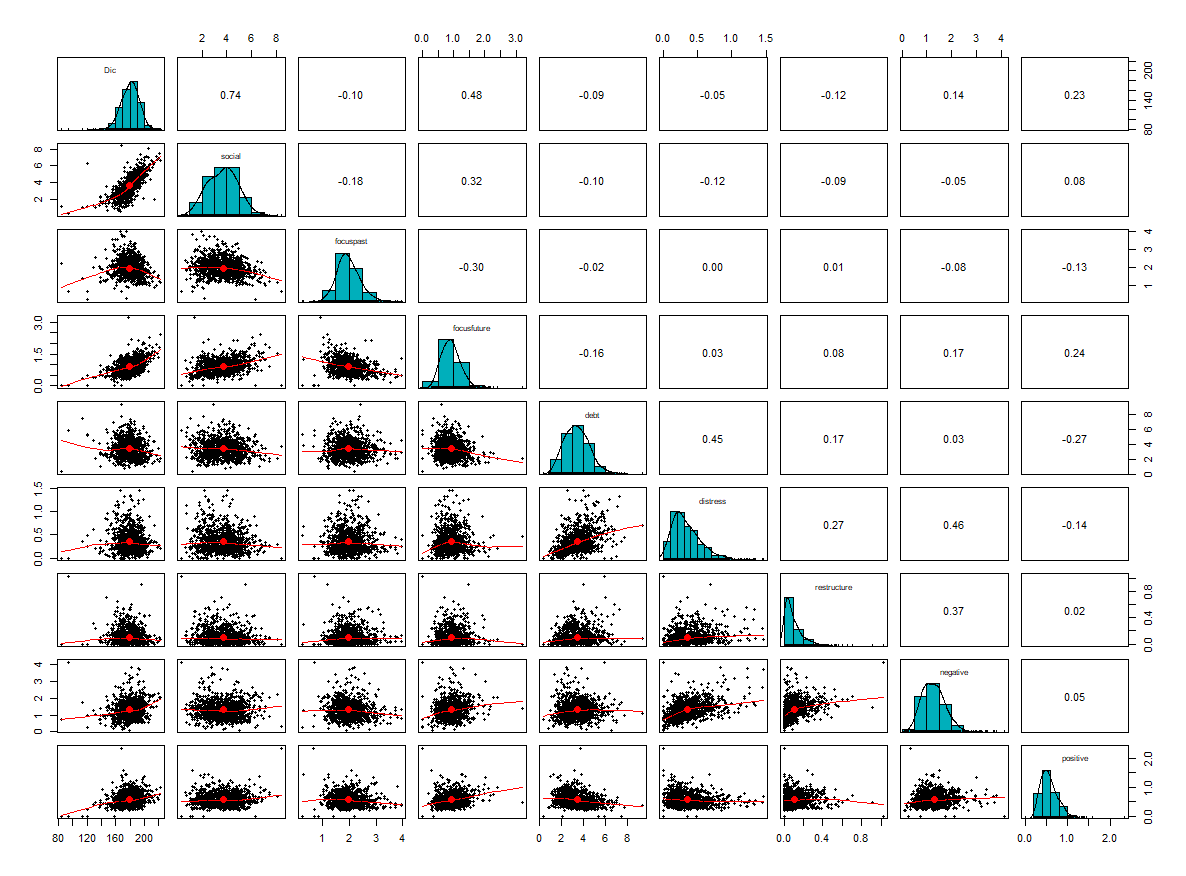} \caption{Correlations between all selected linguistic features}\label{fig:LinguisticCorrelations}
\end{figure}
\end{landscape}

\newpage

\hypertarget{experiment-results}{%
\subsection{Experiment results}\label{experiment-results}}

The following will explain the results of various experiments done to
test the hypothesis we outlined in the methodology

\hypertarget{hypothesis-1-linguistic-differences-exist-between-bankrupt-and-non-bankrupt-firms-financial-disclosures}{%
\subsection{Hypothesis 1: Linguistic differences exist between bankrupt
and non-bankrupt firm's financial
disclosures}\label{hypothesis-1-linguistic-differences-exist-between-bankrupt-and-non-bankrupt-firms-financial-disclosures}}

From descriptive statistics, we observed that there are distinct
qualities that differentiate bankrupt firms from non-bankrupt firms. We
set out to test this hypothesis.

\hypertarget{association-between-linguistic-markers-and-distress}{%
\subsubsection{Association between linguistic markers and
distress}\label{association-between-linguistic-markers-and-distress}}

Independent T-tests were conducted. The number of bankrupt firms and
non-bankrupt firms is 500 each

The 500 bankrupt firms compared to the 500 non-bankrupt firms
demonstrated significantly higher distress, t(868) = 17.38, p = .00.

Bankrupt firms had significantly higher debt (t(992) = 12.32, p= 0.00),
higher negative words (t(997) = 8.28, p= 0.00) and higher restructure
words (t(922) = 7.67, p= 0.000)

There was no significant effect for negative emotions (negemo), t(988) =
0.69, p = .62, despite bankrupt (M = 0.88, SD = 0.38) attaining higher
scores than non-bankrupt (M = 0.86, SD = 0.35).

\begin{figure}
\includegraphics[width=1\linewidth]{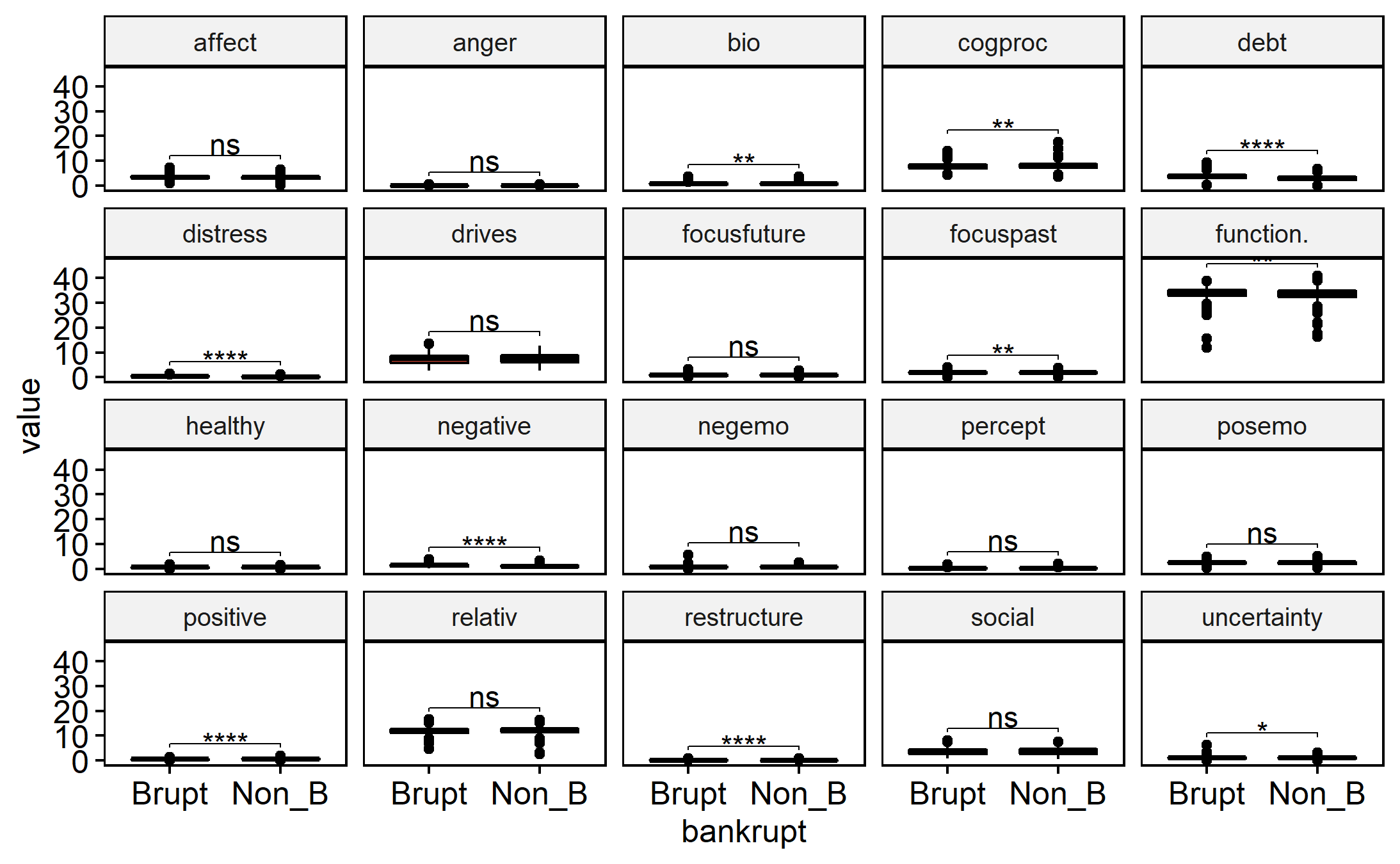} \caption{Bankrupt vs non-bankrupt linguistic features T test}\label{fig:LinguisticTTests}
\end{figure}

\ref{fig:LinguisticTTests} shows the details.

\newpage

\newpage

\hypertarget{hypothesis-2-domain-specific-dictionaries-capture-linguistic-differences-better-than-general-language-models}{%
\subsection{Hypothesis 2: Domain-specific dictionaries capture
linguistic differences better than general language
models}\label{hypothesis-2-domain-specific-dictionaries-capture-linguistic-differences-better-than-general-language-models}}

As part of this hypothesis, a Logistic regression model with all LIWC
features as independent variables has been fit. Another model with L.M.
features as predictors are built and compared.

\hypertarget{liwc-1}{%
\subsubsection{LIWC}\label{liwc-1}}

Here we review the LIWC logit model. Table \ref{tab:LIWCLogit} presents
the model details.

We can observe that only a few predictors are significant. This
observation is expected as the LIWC model captures various aspects of
language, and only a few of them can be expected to be impacted by the
distress and potential bankruptcy conditions. \(WPS\), \(Dic\),
\(function.\), \(focuspast\), and \(focusfuture\) are significant at
0.001 level. The logistic regression coefficients give the change in the
log odds of the outcome for a one-unit increase in the predictor
variable. Here, except \(WPS\), all predictors are percentages of
category words.

For every one unit change in \(WPS\), the log odds of bankruptcy (versus
non-bankruptcy) increases by 0.08 with 95\% CI {[}0.04, 0.12{]}. For a
one percent increase in \(focuspast\) the log odds of being bankrupt
increases by 1.10 with 95\% CI {[}0.66, 1.55{]}. The same for
\(focusfuture\) increases by 1.65 with 95\% CI {[}1.00, 2.31{]}.

Another way to interpret these coefficients is to use the odds ratio.
This fitted model says that holding other predictors at a fixed value,
the odds of bankruptcy for a firm whose disclosure has 1\%
\(focusfuture\) words than a firm with zero percent such words are
exp(1.65) = 5.2. We can say that the odds for a firm with \(1%
\) higher \(focusfuture\) words are 420\% higher in terms of percent
change.

Other predictors that are significant at \textless0.05 levels are
\(social\), \(cogproc\), and \(drives\). The log odds are 0.37, -0.34,
0.30 with 95\% CIs {[}0.10, 0.65{]}, {[}-0.64, -0.06{]}, and {[}-0.00,
-0.60{]}, respectively.

\newpage
\begin{table}

\caption[LIWC model coefficients]{\label{tab:LIWCLogit}LIWC model coefficients }
\centering
\begin{tabular}[t]{l|r|r|l|r|r|r}
\hline
Predictors & Coefficients & SE & pvalue & Lower CI & Upper CI & Odds Ratio\\
\hline
Intercept & 2.75 & 3.34 & 0.411 & -3.92 & 9.16 & 15.68\\
\hline
WPS & 0.08 & 0.02 & <0.001 & 0.04 & 0.12 & 1.08\\
\hline
WC & 0.00 & 0.00 & 0.103 & 0.00 & 0.00 & 1.00\\
\hline
Sixltr & -0.07 & 0.05 & 0.202 & -0.17 & 0.04 & 0.93\\
\hline
Dic & -0.16 & 0.04 & <0.001 & -0.24 & -0.08 & 0.85\\
\hline
function. & 0.60 & 0.11 & <0.001 & 0.39 & 0.81 & 1.82\\
\hline
affect & -0.60 & 5.21 & 0.909 & -11.50 & 9.26 & 0.55\\
\hline
social & 0.37 & 0.14 & 0.007 & 0.10 & 0.65 & 1.45\\
\hline
cogproc & -0.34 & 0.15 & 0.021 & -0.64 & -0.06 & 0.71\\
\hline
percept & 0.73 & 0.40 & 0.070 & -0.05 & 1.52 & 2.07\\
\hline
bio & -0.05 & 0.23 & 0.829 & -0.49 & 0.39 & 0.95\\
\hline
drives & 0.30 & 0.15 & 0.049 & 0.00 & 0.60 & 1.35\\
\hline
relativ & -0.06 & 0.12 & 0.604 & -0.29 & 0.17 & 0.94\\
\hline
AllPunc & -0.08 & 0.04 & 0.062 & -0.16 & 0.00 & 0.92\\
\hline
focuspast & 1.10 & 0.23 & <0.001 & 0.66 & 1.55 & 3.00\\
\hline
focuspresent & 0.06 & 0.21 & 0.773 & -0.35 & 0.48 & 1.06\\
\hline
focusfuture & 1.65 & 0.33 & <0.001 & 1.00 & 2.31 & 5.20\\
\hline
anger & 0.68 & 2.18 & 0.754 & -3.57 & 5.05 & 1.98\\
\hline
posemo & 1.15 & 5.21 & 0.825 & -8.70 & 12.05 & 3.17\\
\hline
negemo & 1.34 & 5.24 & 0.799 & -8.57 & 12.30 & 3.81\\
\hline
\end{tabular}
\end{table}

\begin{figure}
\includegraphics[width=1\linewidth]{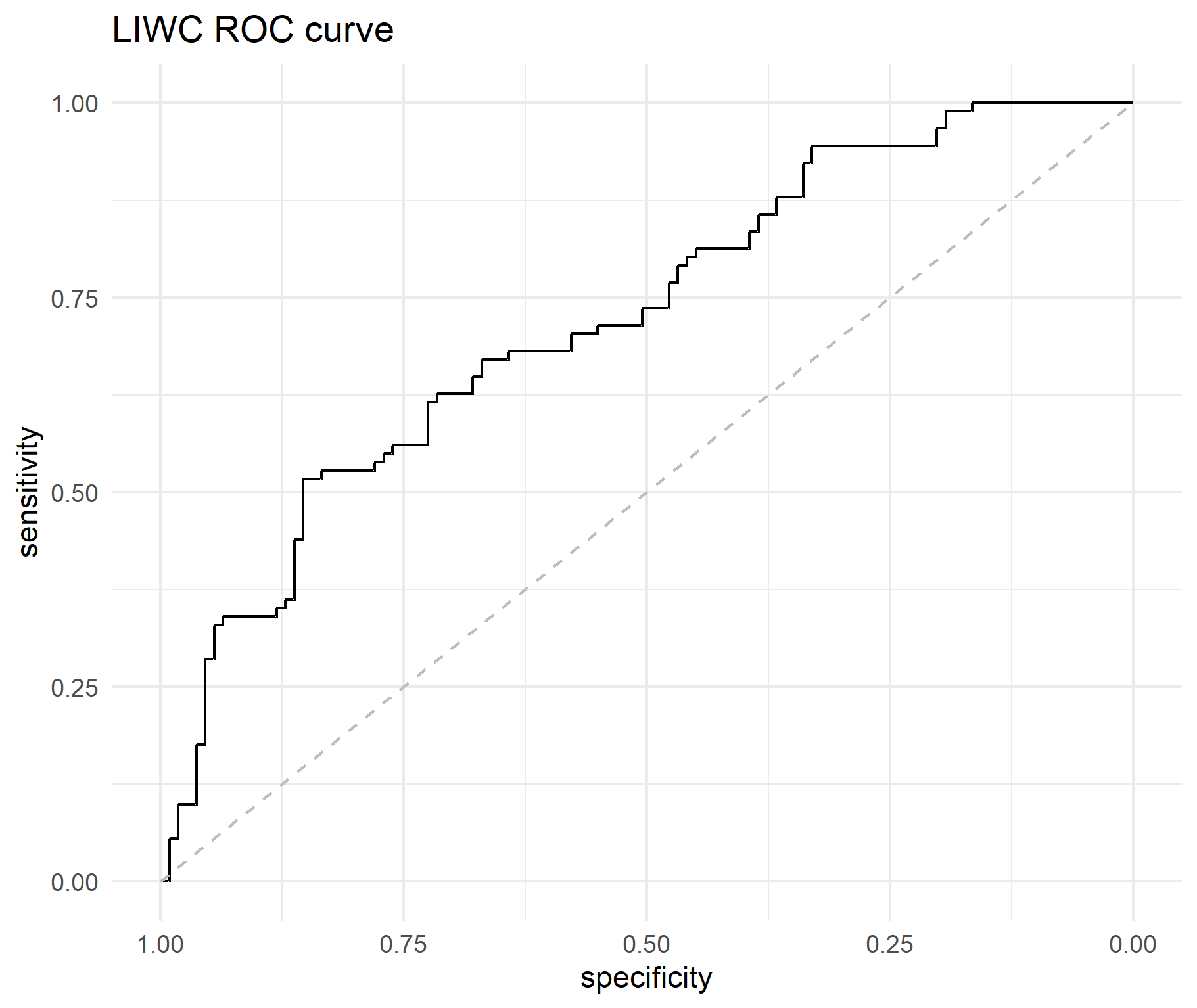} \caption{LIWC model ROC}\label{fig:LIWCROC}
\end{figure}

\newpage

\hypertarget{lm}{%
\subsubsection{LM}\label{lm}}

Here we review the L.M. logit model. Table \ref{tab:LMLogit} presents
the coefficients and confidence intervals.

We can observe that all predictors are significant. \(negative\) and
\(positive\) are significant at 0.001 level. For a one percent increase
in \(negative\) words, the log odds of being bankrupt increases by 1.141
with 95\% CI {[}1.08, 1.76{]}. The same for \(positive\) changes by
-2.88 with 95\% CI {[}-3.68, -2.12{]}.

This L.M. model says that holding other predictors at a fixed value, the
odds of bankruptcy for firms whose disclosure has 1\% \(negative\) words
than a firm with zero percent such words are exp(1.41) = 4.10. We can
say that the odds for a firm with \(1%
\) higher \(negative\) words are 310\% higher in terms of percent
change.

\begin{table}

\caption[LM model coefficients]{\label{tab:LMLogit}LM model coefficients }
\centering
\begin{tabular}[t]{l|r|r|l|r|r|r}
\hline
Predictors & Coefficients & SE & pvalue & Lower CI & Upper CI & Odds Ratio\\
\hline
Intercept & 0.56 & 0.31 & 0.070 & -0.04 & 1.17 & 1.75\\
\hline
negative & 1.41 & 0.17 & <0.001 & 1.08 & 1.76 & 4.10\\
\hline
positive & -2.88 & 0.40 & <0.001 & -3.68 & -2.12 & 0.06\\
\hline
uncertainty & -0.64 & 0.20 & 0.002 & -1.06 & -0.26 & 0.53\\
\hline
\end{tabular}
\end{table}

\begin{figure}
\includegraphics[width=1\linewidth]{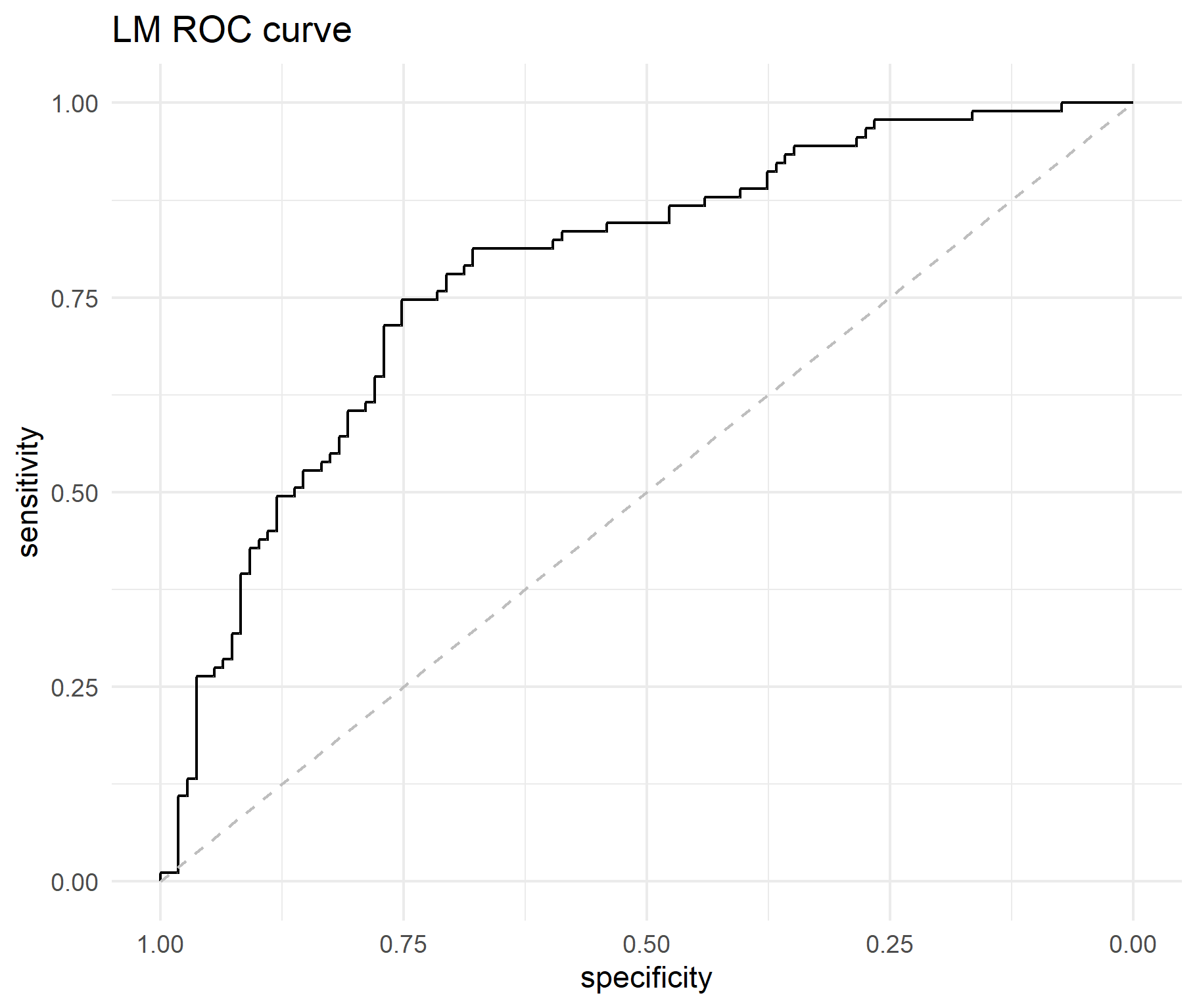} \caption{LM model ROC}\label{fig:LMROC}
\end{figure}

\hypertarget{liwc-and-l.m.-model-comparison}{%
\subsubsection{LIWC and L.M. model
comparison}\label{liwc-and-l.m.-model-comparison}}

ANOVA indicates that models are significantly different.

Accuracy, BIC, and AIC metrics

\begin{table}

\caption[LIWC and LM model comparison]{\label{tab:LIWCLMCompare}LIWC and LM model comparison }
\centering
\begin{tabular}[t]{lrrrrrrrr}
\toprule
Model & Training Accuracy & Test Accuracy & LogLik & AIC & BIC & AUC & Deviance & Parameters\\
\midrule
LIWC & 0.69 & 0.63 & -470.61 & 981.22 & 1074.92 & 0.72 & 941.22 & 19\\
LM & 0.68 & 0.74 & -489.88 & 987.75 & 1006.49 & 0.78 & 979.75 & 3\\
\bottomrule
\end{tabular}
\end{table}

\begin{figure}
\includegraphics[width=1\linewidth]{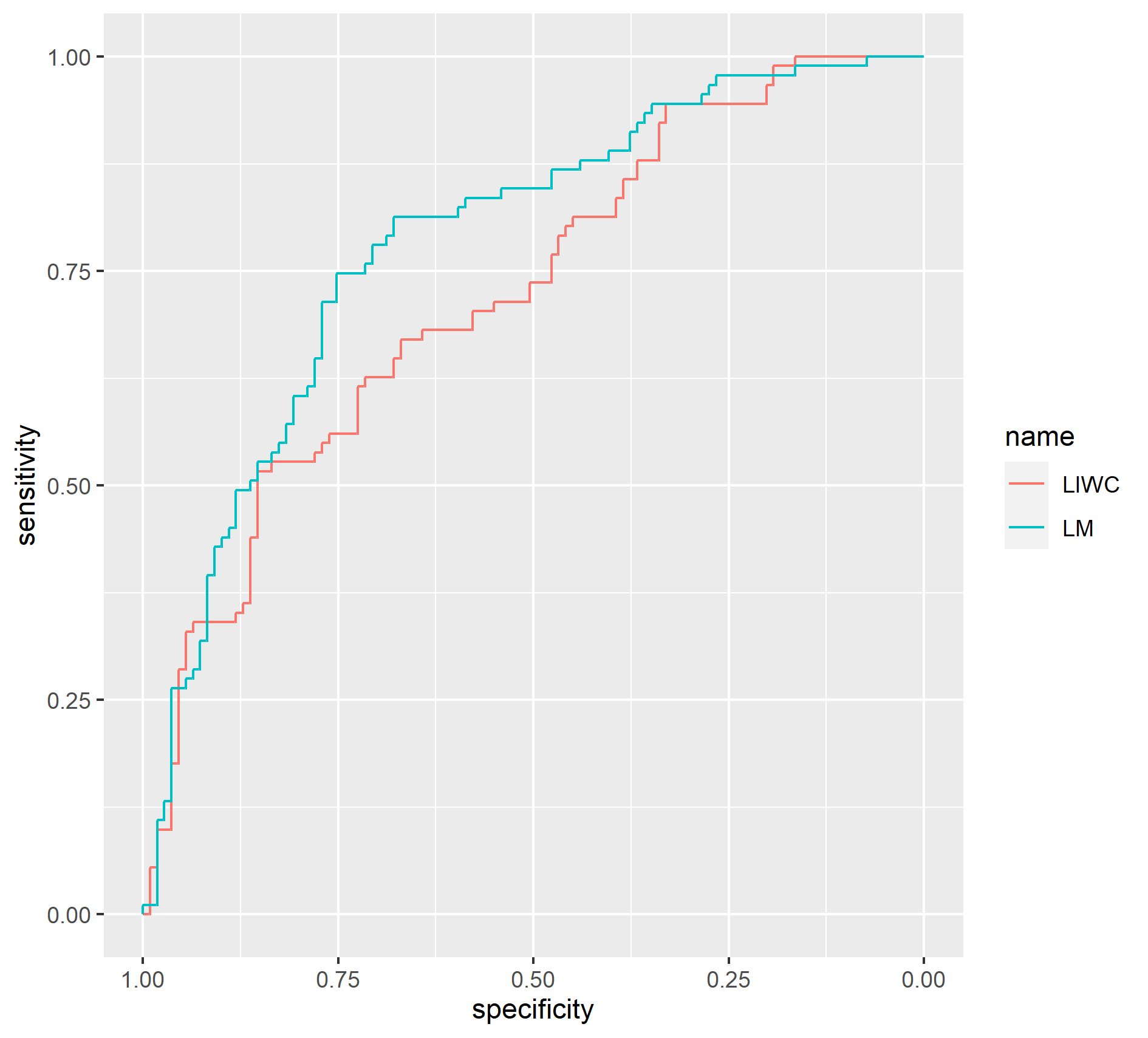} \caption{LIWC LM ROC comparison}\label{fig:LIWCLMROC}
\end{figure}

ROC comparison is shown in figure \ref{fig:LIWCLMROC}

We can observe that for the L.M. model, while BIC is lower than the LIWC
model, AIC is higher. Recall that we noted in section \ref{BIC}, for
sample size \textgreater100, BIC will prefer smaller models for similar
log-likelihoods. The out of sample forecasting performance represented
in the \emph{Test Accuracy} column indicates the L.M. model provides
10\% higher accuracy. Also, ROC is better for L.M. Overall, while the
LIWC model captures more information, probably due to many parameters,
the L.M. model predictive performance is better than the LIWC model.

\newpage

\hypertarget{hypothesis-3-task-specific-dictionaries-capture-linguistic-differences-better-than-domain-specific-dictionaries}{%
\subsection{Hypothesis 3: Task-specific dictionaries capture linguistic
differences better than domain-specific
dictionaries}\label{hypothesis-3-task-specific-dictionaries-capture-linguistic-differences-better-than-domain-specific-dictionaries}}

\hypertarget{stress-dictionary-1}{%
\subsubsection{Stress dictionary}\label{stress-dictionary-1}}

Here we review the stress dictionary logit model. Table
\ref{tab:StressScoreLogit} presents the coefficients and confidence
intervals.

We observe that \(debt\), \(distress\) and \(restructure\) are
significant at 0.001 level. For a one percent increase in \(distress\)
words, the log odds of being bankrupt increases by 5.03 with 95\% CI
{[}3.98, 6.15{]}. The same for \(debt\), \(restructure\) increase by
0.36 and 2.96 with 95\% CIs {[}0.19, 0.54{]} and {[}1.45, 4.54{]}
respectively.

Most importantly, as per this model, holding other predictors at a fixed
value, the odds of bankruptcy for a firm whose disclosure has 1\%
\(distress\) words compared to a firm with zero percent such words is
exp(5.03) = 153.66. This high odds ratio indicates that distress words
percentage is a highly sensitive indicator to forthcoming bankruptcy.

\begin{table}

\caption[Stress model coefficients]{\label{tab:StressScoreLogit}Stress dictionary model coefficients }
\centering
\begin{tabular}[t]{l|r|r|l|r|r|r}
\hline
Predictors & Coefficients & SE & pvalue & Lower CI & Upper CI & Odds Ratio\\
\hline
Intercept & -3.36 & 0.40 & <0.001 & -4.16 & -2.59 & 0.03\\
\hline
debt & 0.36 & 0.09 & <0.001 & 0.19 & 0.54 & 1.44\\
\hline
distress & 5.03 & 0.55 & <0.001 & 3.98 & 6.15 & 153.66\\
\hline
restructure & 2.96 & 0.79 & <0.001 & 1.45 & 4.54 & 19.39\\
\hline
healthy & 0.23 & 0.38 & 0.5 & -0.51 & 0.98 & 1.26\\
\hline
\end{tabular}
\end{table}

\begin{figure}
\includegraphics[width=1\linewidth]{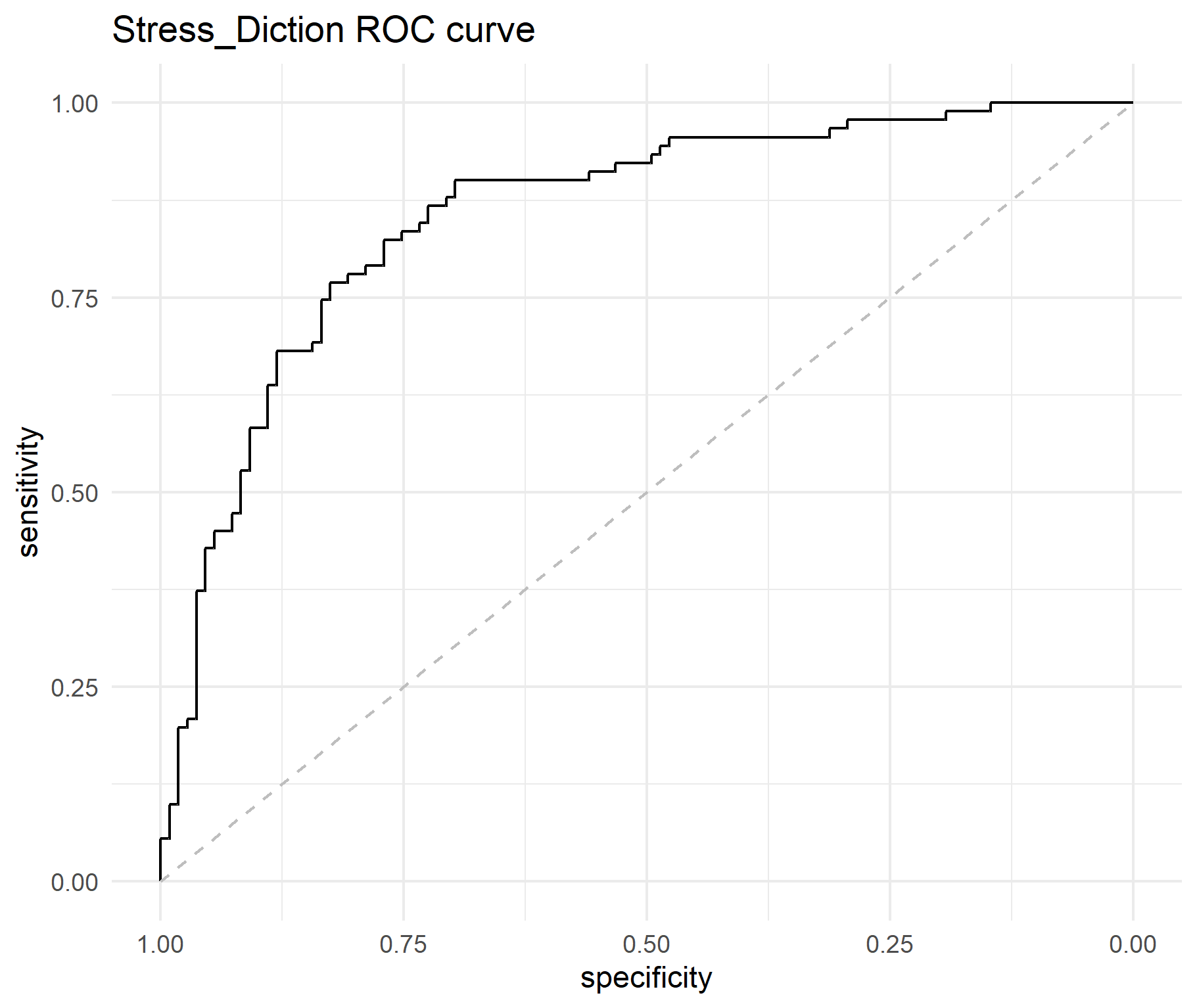} \caption{Stress dictionary model ROC}\label{fig:StressScoreROC}
\end{figure}

\hypertarget{stress-dictionary-vs.-l.m.}{%
\subsubsection{Stress dictionary
vs.~L.M.}\label{stress-dictionary-vs.-l.m.}}

\begin{table}

\caption[Stress and LM model comparison]{\label{tab:StressLMCompare}Stress and LM model comparison }
\centering
\begin{tabular}[t]{lrrrrrrrr}
\toprule
Model & Training Accuracy & Test Accuracy & LogLik & AIC & BIC & AUC & Deviance & Parameters\\
\midrule
LM & 0.68 & 0.74 & -489.88 & 987.75 & 1006.49 & 0.78 & 979.75 & 3\\
Stress\_Diction & 0.72 & 0.79 & -428.09 & 866.18 & 889.61 & 0.86 & 856.18 & 4\\
\bottomrule
\end{tabular}
\end{table}

\begin{figure}
\includegraphics[width=1\linewidth]{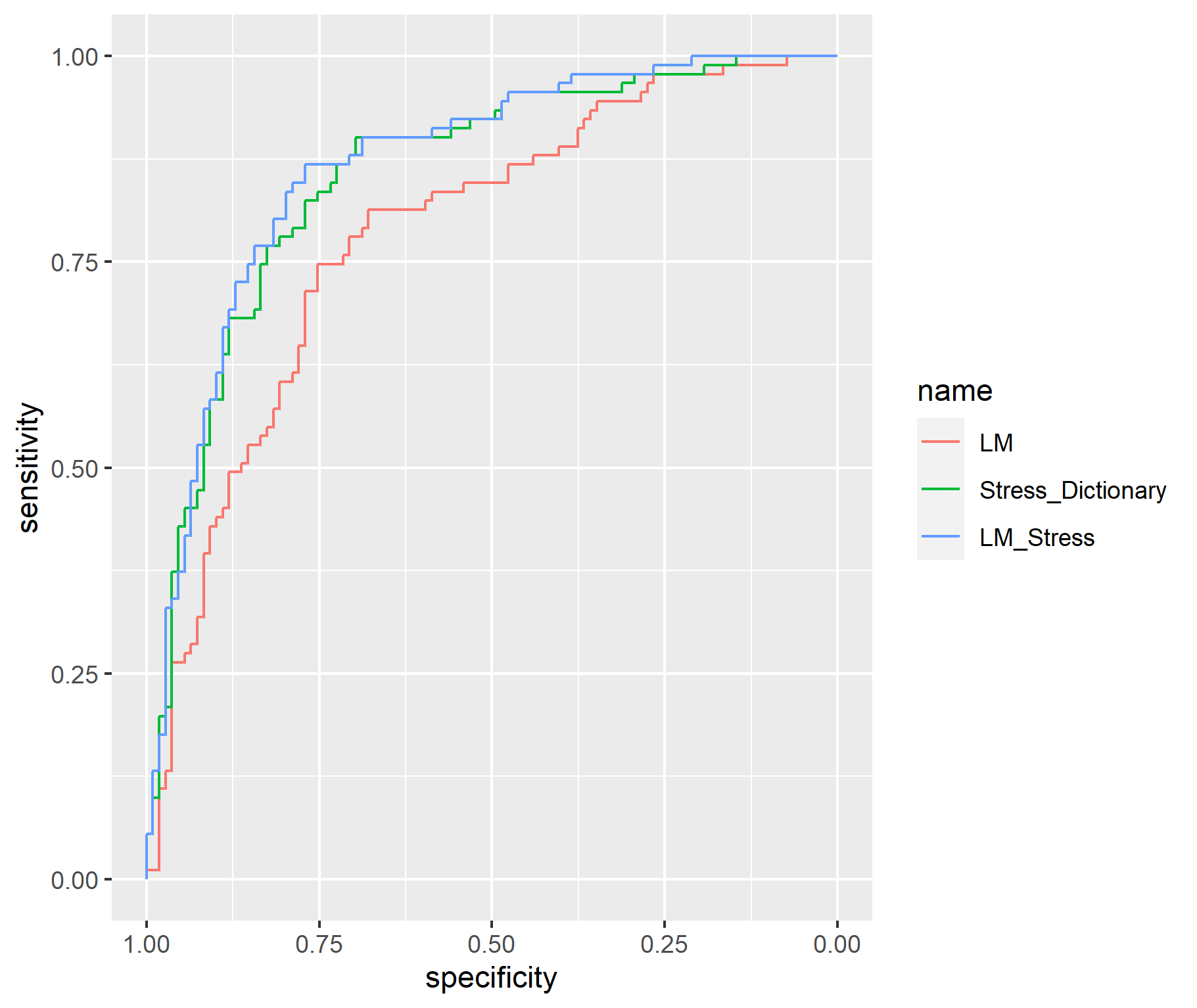} \caption{LM and stress dictionary model ROC comparison}\label{fig:LMStressScoreROCCompare}
\end{figure}

ROC comparison is shown in figure \ref{fig:LMStressScoreROCCompare}
Overall, we can observe that the Stress model is better than the L.M.
model on BIC and ROC criteria. Also, test performance is better in the
Stress dictionary.

\hypertarget{hypothesis-3.1-combination-models-outperform-individual-dictionary-models}{%
\subsubsection{Hypothesis 3.1: Combination models outperform individual
dictionary
models}\label{hypothesis-3.1-combination-models-outperform-individual-dictionary-models}}

Considering the observation that the Correlation between LIWC, L.M., and
stress dictionary features is low, we can take advantage of their
complementary nature. Three combination models with combined inputs have
been fitted on the dataset: LIWC + Stress, L.M. + Stress, and LIWC +
L.M. + Stress. Model coefficients are presented in appendix B. The
performance results are shared below.

\newpage

\hypertarget{combination-models}{%
\paragraph{Combination models}\label{combination-models}}

\begin{table}

\caption[Dictionary models AUC comparison]{\label{tab:HybridLogit}Dictionary models AUC comparison }
\centering
\begin{tabular}[t]{lr}
\toprule
Model & AUC\\
\midrule
LIWC & 0.72\\
LM & 0.78\\
Stress\_Diction & 0.86\\
LIWC\_Stress & 0.86\\
LM\_Stress & 0.87\\
\addlinespace
LIWC\_LM\_Stress & 0.87\\
\bottomrule
\end{tabular}
\end{table}

\begin{figure}
\includegraphics[width=1\linewidth]{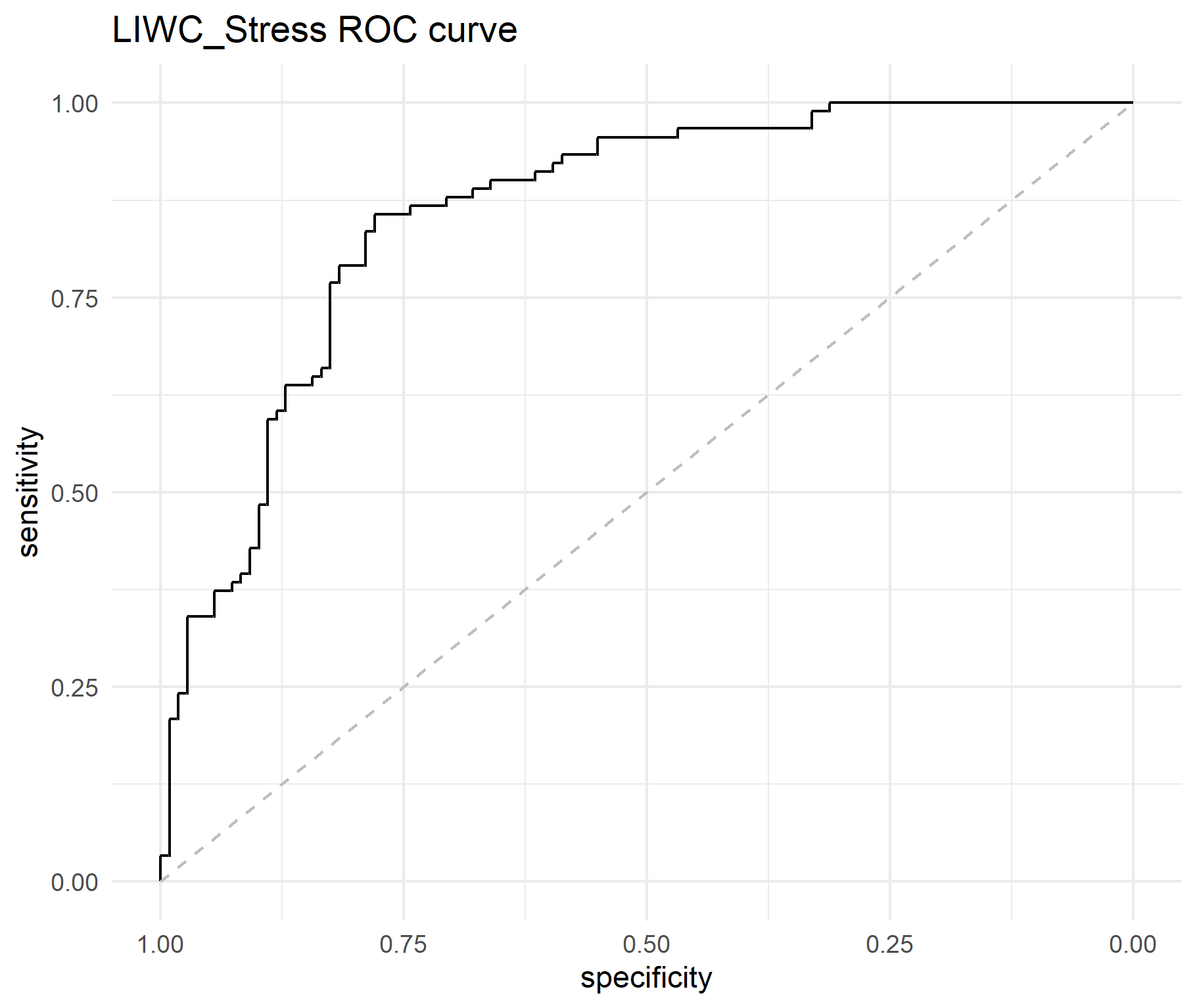} \caption{LIWC and stress dictionary model ROC}\label{fig:LIWCStressScoreROC}
\end{figure}

\begin{figure}
\includegraphics[width=1\linewidth]{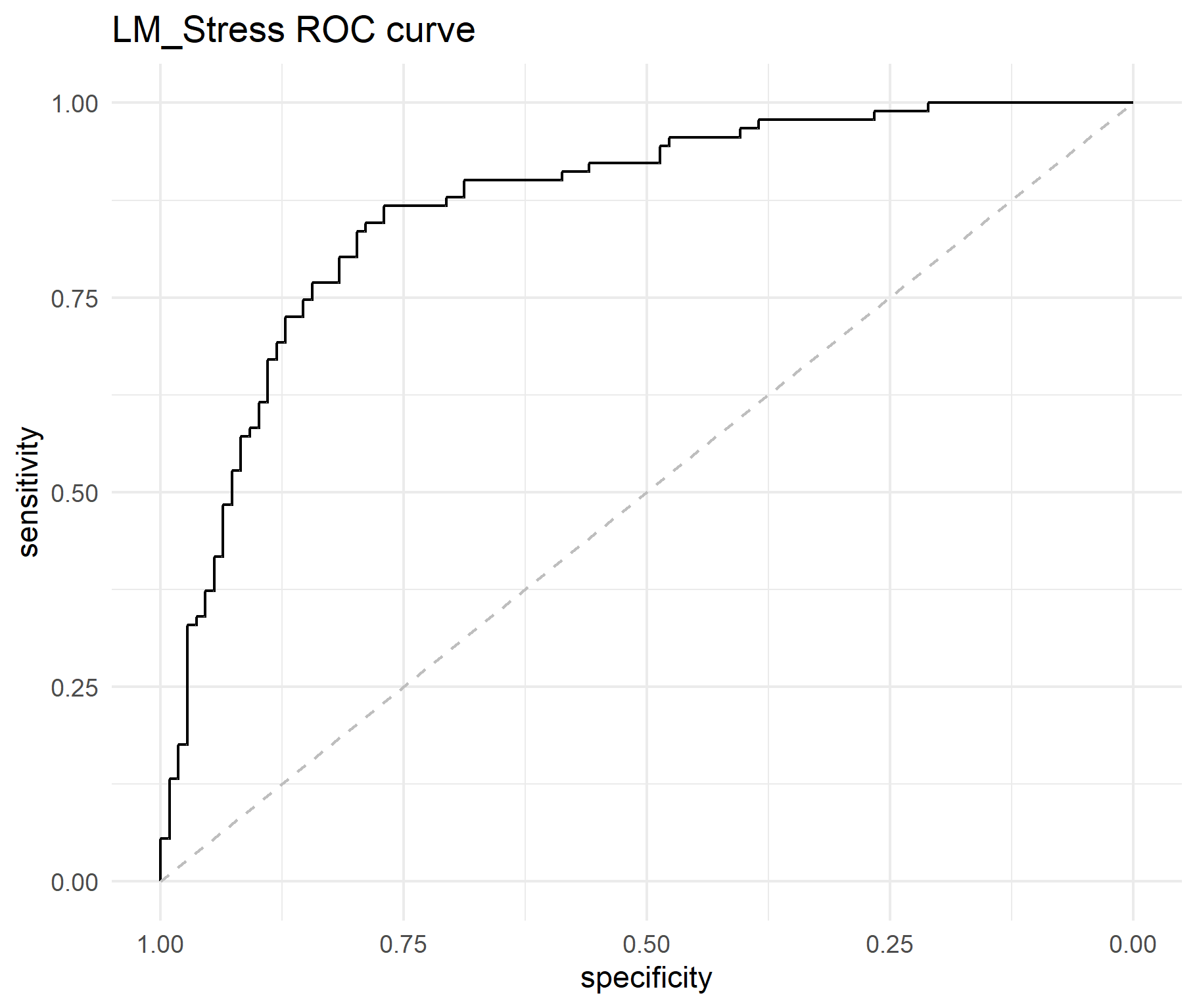} \caption{LM and stress dictionary model ROC}\label{fig:LMStressScoreROC}
\end{figure}

\begin{figure}
\includegraphics[width=1\linewidth]{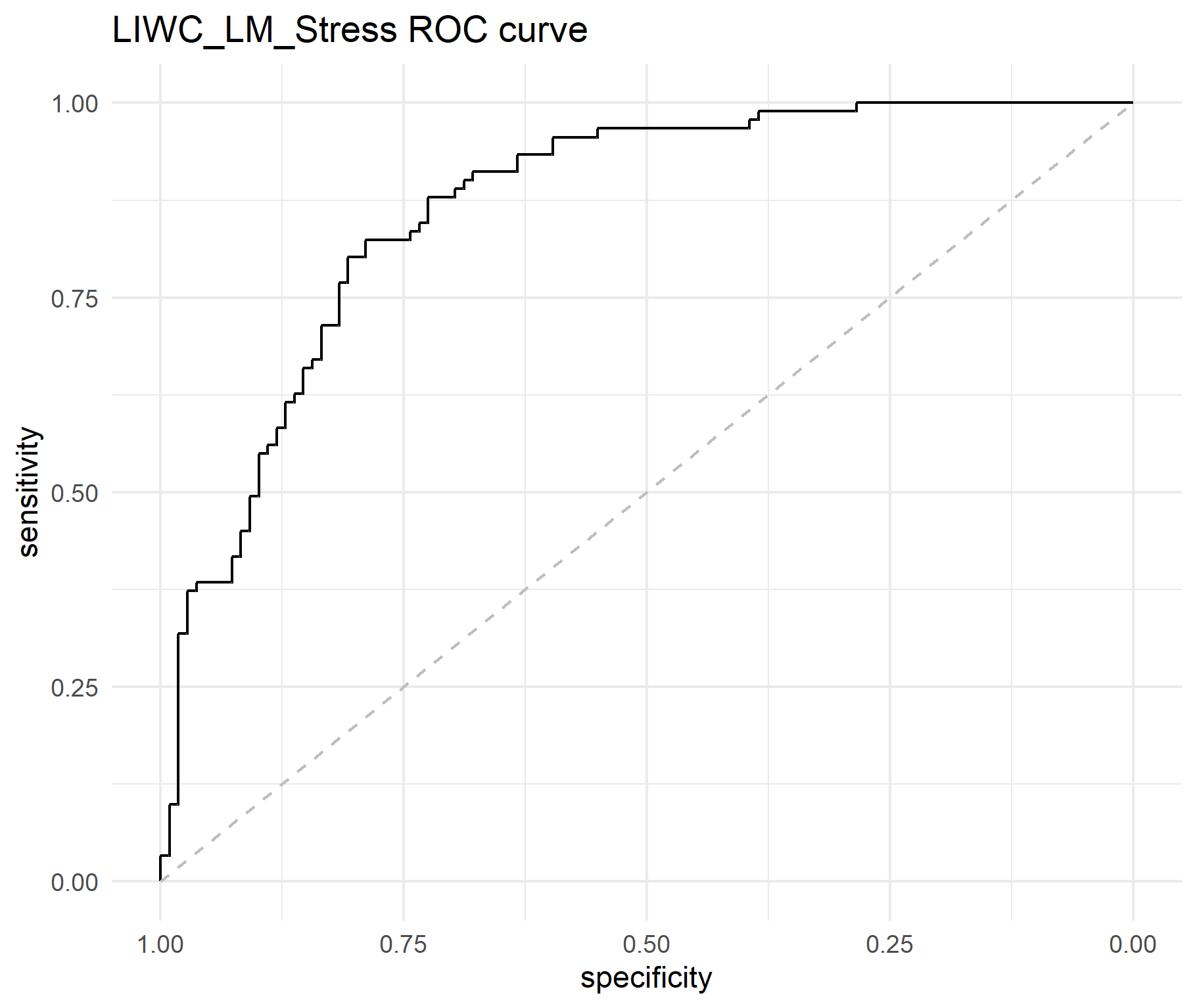} \caption{LIWC, LM and stress dictionary model ROC comparison}\label{fig:LIWCLMStressScoreROC}
\end{figure}

\newpage

\hypertarget{summary-of-dictionary-based-models}{%
\subsubsection{Summary of dictionary-based
models}\label{summary-of-dictionary-based-models}}

This subsection reviews the dictionary-based models. We have evidence to
believe that there is incremental performance improvement as additional
features are incorporated into the model. A comparison of model
performance is as below

\begin{table}

\caption[Dictionary models comparison]{\label{tab:BOWcomparison}Dictionary models comparison }
\centering
\begin{tabular}[t]{lrrrrrrrr}
\toprule
Model & Training Accuracy & Test Accuracy & LogLik & AIC & BIC & AUC & Deviance & Parameters\\
\midrule
LIWC\_LM\_Stress & 0.78 & 0.80 & -366.95 & 787.89 & 914.38 & 0.87 & 733.89 & 26\\
LIWC\_Stress & 0.77 & 0.79 & -374.89 & 797.78 & 910.21 & 0.86 & 749.78 & 23\\
LM\_Stress & 0.74 & 0.80 & -410.58 & 837.17 & 874.64 & 0.87 & 821.17 & 7\\
Stress\_Diction & 0.72 & 0.79 & -428.09 & 866.18 & 889.61 & 0.86 & 856.18 & 4\\
LIWC & 0.69 & 0.63 & -470.61 & 981.22 & 1074.92 & 0.72 & 941.22 & 19\\
\addlinespace
LM & 0.68 & 0.74 & -489.88 & 987.75 & 1006.49 & 0.78 & 979.75 & 3\\
\bottomrule
\end{tabular}
\end{table}

\begin{figure}
\includegraphics[width=1\linewidth]{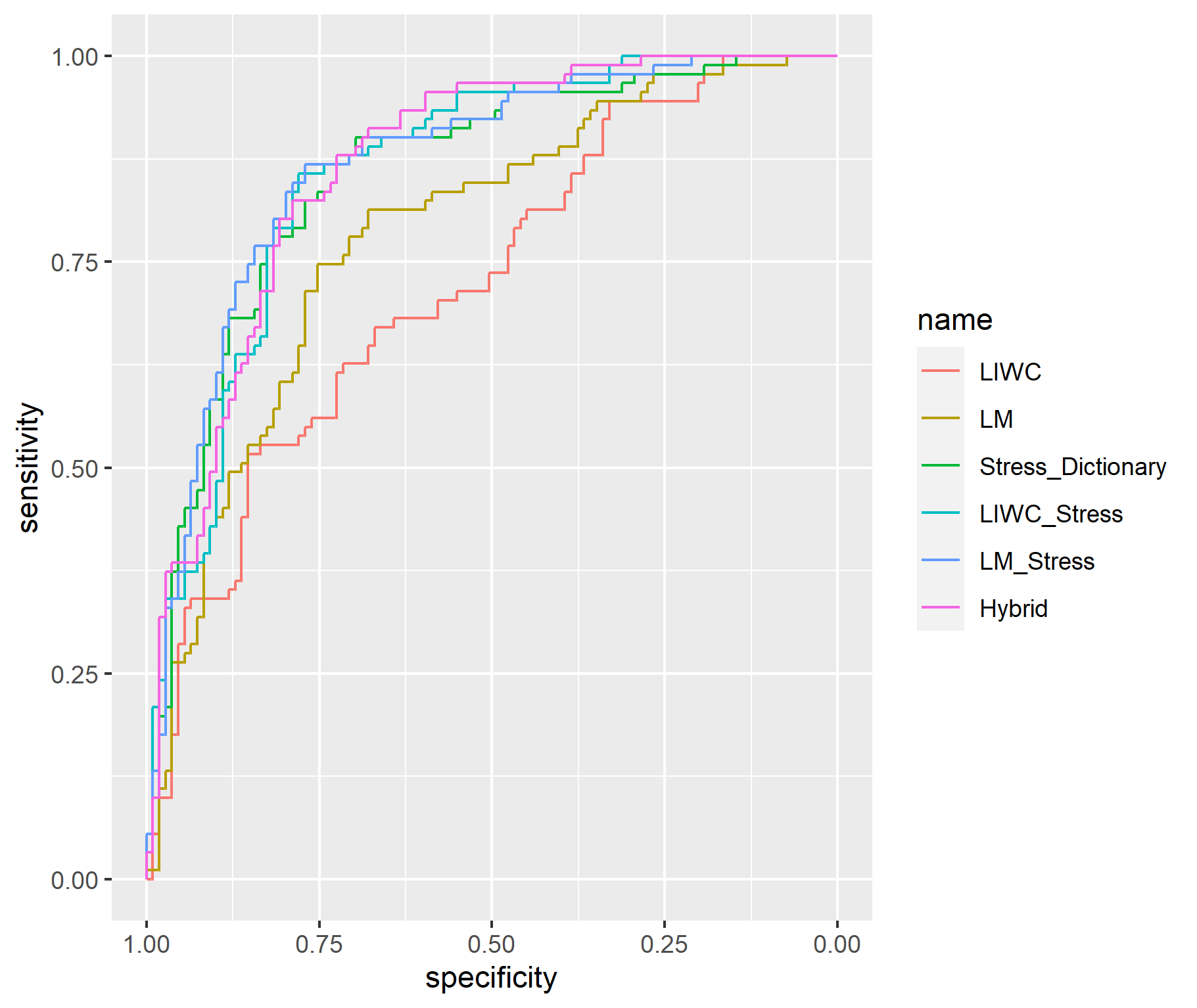} \caption{Bag of words dictionary models ROC comparison}\label{fig:BOWROCBoost}
\end{figure}

\hypertarget{discussion}{%
\section{Discussion}\label{discussion}}

This work provides the first comprehensive test of text disclosure-based
dictionary-based bankruptcy prediction models. For Dictionary-based
models, I apply the LIWC dictionary Pennebaker et al. (2015), the
Loghron McDonalds Dictionary Loughran and Mcdonald (2011), and a custom
dictionary, developed as part of this work. To test the models'
performance, I use receiver operating characteristics (ROC) curves,
information content tests, and the accuracy metrics.

The tests using ROC curve analysis demonstrated that all
dictionary-based bankruptcy prediction models have a greater forecasting
accuracy than a random model and that the composite models perform
better than their individual language models. Information content tests
provide evidence that all models carry significant bankruptcy-related
information.

\hypertarget{Conclusions}{%
\section{Conclusion}\label{Conclusions}}

In this chapter, I summarize the contributions of the current study to
text analysis in finance, present the research's objectives and findings
in the context of previous research, and suggest appropriate future
research directions.

\hypertarget{objectives-and-summary-of-research}{%
\subsection{Objectives and summary of
research}\label{objectives-and-summary-of-research}}

This study constitutes an exploration of knowledge extraction from the
narrative, corporate report sections using text analysis. The aims are

1.To establish if linguistic features of disclosures can explain firm
attributes in the financial analysis context\\
2.To determine which language models perform better in capturing
information.\\
3.Specifically, to predict bankruptcy based on management's discussion
and analysis in annual filings.

Knowledge in a public firm's context involves information that can
influence organizational outcomes and future stock performance. As
managers have an information advantage over the public, their narrative
disclosures have significant information content, over and above the
quantitative financial measures. This information helps in understanding
the firm's current financial status, the firm's ability to continue its
operations without hindrances, the kinds of risks the firm is exposed
to, strategic and tactical interventions the management is undertaking
to overcome the challenges capture the opportunities and capital
allocation plans. This knowledge gives more in-depth insights into the
firm's prospects.

In the context of this thesis, knowledge extraction is studied in the
form of predicting adverse organizational outcomes, specifically in the
form of (1) Bankruptcy prediction, i.e., predict if a firm will file for
chapter 7 or 11 within one year after the annual filing date (2) using
management disclosures and analysis section in annual filing (10-K).

For this purpose, I employed one new measurement technique based on
content analysis and research, namely a stress score based on the number
of financial stress words per thousand words. I built text feature-based
bankruptcy prediction models LIWC dictionary, L.M. dictionary, and
Stress dictionary. Concerning the text feature-based bankruptcy
prediction introduced in this study, this is the first time they are
used in accounting research, and they address the numbers bias concerns
inherent in traditional approaches. The scoring of Stress using
linguistic markers is also a new approach to measuring financial
distress. It is based on the linguistic characteristics that management
displays when explaining the current liquidity challenges and its
attempts to overcome them through debt extensions, new financing, and
asset restructuring. Such explanations would result in increased
frequencies of words related to covenants, modified loan agreements,
restructuring, new financing activities, uncertainty about firms'
ability to raise funds, asset sales, and capital expense reduction in a
distressed corporate reporting context. It may also result in management
attempting to present a rosy image of prospects to outsiders
inconsistent with management's perception of the firm and its
performance.

\hypertarget{summary-of-result-and-comparison-with-previous-research-findings}{%
\subsection{Summary of result and comparison with previous research
findings}\label{summary-of-result-and-comparison-with-previous-research-findings}}

Bankruptcy prediction has been an active research topic for accounting
researchers over decades. With the improved awareness about financial
ratios' shortcomings and availability of text analysis tools,
researchers have explored incorporating textual features into bankruptcy
prediction models. Hájek and Olej (2015) studied various word categories
from corporate annual reports and showed that the language used by
bankrupt companies shows stronger tenacity, accomplishment, familiarity,
present concern, exclusion, and denial. They built prediction models
combining both financial indicators and word categorizations as input
variables.

Working on U.S. Banks Gandhi, Loughran, and McDonald (2017) used
disclosure text sentiment as a proxy for bank distress. Other notable
works using text analysis for bankruptcy prediction were Yang, Dolar,
and Mo (2018) and Mayew, Sethuraman, and Venkatachalam (2015). Yang,
Dolar, and Mo (2018) used high-frequency words from MDA and compared the
differences between bankrupt and non-bankrupt companies. Mayew,
Sethuraman, and Venkatachalam (2015) also analyzed MDA with a focus on
going-concern options. They found that disclosure's predictive ability
is incremental to financial ratios, market-based variables, even the
auditor's going concern opinion and extends to three years before the
bankruptcy.\\
As we can observe, prior work focused on marginal information content in
the text. While researchers concluded that narratives have information
content and predictive power, the limits and extent of that information
are not tested. This thesis tests that and demonstrates that the
information content is sufficient to predict bankruptcy, independent of
any financial and quantitative metrics.

Prior work in disclosure text analysis focused on simple text measures
like readability, sentiment, and tone. This limitation was probably
motivated by the intent to use them as marginal predictors, along with
financial ratios. Also, the language model methods were the limitation.
Limited organizational outcomes can be explained by shallow language
models that capture marginal information from disclosure text. My work
has demonstrated that interpretable and accurate predictions can be made
with task-specific dictionaries.

\hypertarget{implications-of-research-findings}{%
\subsection{Implications of research
findings}\label{implications-of-research-findings}}

This work demonstrates that textual disclosures, independent of
financial ratios, have predictive power. Further, by way of
task-independent language models, this work enables multiple tasks to be
solved with the same set of features, i.e., language features. With a
sufficiently large dataset containing 100s of samples, researchers can
build reliable predictive models quickly.

Another implication is text-based soft metrics. Investors are interested
in knowing firm performance on corporate responsibility, climate change,
and ethical business practices. It is not easy to measure these
attributes using accounting metrics. Capturing and reporting new metrics
will involve significant capital expenditure for firms. Firms can report
the same using narrative disclosures. Investors can extract the same
using the methods shown in this work. Finally, text metrics-based
factors and factor investment is a possibility based on this approach.
Lopez Lira (2019) used text-based analysis to measure firm risk exposure
and built factor models with such risk portfolios. These models explain
cross-section returns suggesting internal validity. Similar portfolios
on other dimensions like fraud, climate exposure, etc., can be explored
using this thesis's approach.

\hypertarget{limitations-of-research}{%
\subsection{Limitations of research}\label{limitations-of-research}}

Like other empirical studies in finance and language processing, the
results presented in this thesis contain some limitations.

Due to resource constraints for downloading, processing, and storing
extensive text data, the Knowledge extraction from Financial disclosures
has been attempted on a single task of Bankruptcy Prediction using
different methodologies. I also restricted the text content to one type
of corporate narrative document (i.e., Management's Discussion and
Analysis). For this reason, caution needs to be applied in generalizing
results.

Since the text analysis is restricted to the surface structure of
language, it is impossible to say whether the extracted signal is a true
reflection of the management's statement. What is more, disclosure
changes can result from managerial interventions, restructuring
activities, e.g., raising capital, asset sell-off or cost reductions,
and analyzing. These interventions can improve performance, and the
distressed firm might show better market performance, hence avoiding
bankruptcy.

I use the EDGAR filing date as the time stamp for filing. If bankruptcy
filing happens within one such filing date, such filing is used to
compute linguistic features, subsequently used as predictors. Any
actions that management takes after filing, which can alter firm stress
levels, are not captured. This limitation is inherent when using
disclosure data.

\hypertarget{suggestions-for-further-research}{%
\subsection{Suggestions for further
research}\label{suggestions-for-further-research}}

The majority of prior text analysis research in the finance context
focuses solely on sentiment analysis and does not address direct
knowledge extraction. In particular, significant effort has been
deployed in linking sentiment and tone to subsequent performance and
fraud. Moreover, the extracted information, i.e., sentiment score, is
used only as an additional input to existing quantitative models. I see
four broad questions that need to be addressed. Given that management
has an information advantage about firms

1.What knowledge can be extracted from the management's textual
disclosures?\\
2.Which of the firm's future states can disclosures textual analysis
explain or predict?\\
3.Which language and document models facilitate fast and reliable
information extraction?\\
4.How can investors incorporate this information into their
decision-making process?

These research questions have received relatively less attention by
comparison with efforts to measure sentiment. Improving the
affordability of data science tools is making unstructured analysis
easier. Wider adoption of unstructured analysis will allow researchers
to apply more focus to these four questions.

\hypertarget{theoretical-perspectives}{%
\subsubsection{Theoretical
perspectives}\label{theoretical-perspectives}}

My work demonstrates that textual disclosures, independent of financial
ratios, have predictive power. This observation raises the question: of
all available financial and accounting metrics, which can be replaced
with more reliable text-based metrics? Text-based metrics can not
possibly possess all the information contained in accounting metrics.
However, it is critical to understand the limits of such information as
well as validity.

\hypertarget{refs}{}
\begin{cslreferences}
\leavevmode\hypertarget{ref-altman1968financial}{}%
Altman, Edward I. 1968. ``Financial Ratios, Discriminant Analysis and
the Prediction of Corporate Bankruptcy.'' \emph{The Journal of Finance}
23 (4): 589--609.

\leavevmode\hypertarget{ref-altman2010corporate}{}%
Altman, Edward I, and Edith Hotchkiss. 2010. \emph{Corporate Financial
Distress and Bankruptcy: Predict and Avoid Bankruptcy, Analyze and
Invest in Distressed Debt}. Vol. 289. John Wiley \& Sons.

\leavevmode\hypertarget{ref-Amel-Zadeh2016}{}%
Amel-Zadeh, Amir, and Jonathan Faasse. 2016. ``The Information Content
of 10-K Narratives: Comparing MD\&A and Footnotes Disclosures.''
\url{https://doi.org/10.2139/ssrn.2807546}.

\leavevmode\hypertarget{ref-Ball2012}{}%
Ball, Christopher, Gerard Hoberg, and Vojislav Maksimovic. 2012.
``Redefining Financial Constraints: A Text-Based Analysis.'' \emph{SSRN
Electronic Journal}. \url{https://doi.org/10.2139/ssrn.1923467}.

\leavevmode\hypertarget{ref-Beams2015}{}%
Beams, Joseph, and Yun Chia Yan. 2015. ``The effect of financial crisis
on auditor conservatism: US evidence.'' \emph{Accounting Research
Journal} 28 (2): 160--71.
\url{https://doi.org/10.1108/ARJ-06-2013-0033}.

\leavevmode\hypertarget{ref-quanteda}{}%
Benoit, Kenneth, Kohei Watanabe, Haiyan Wang, Paul Nulty, Adam Obeng,
Stefan Müller, and Akitaka Matsuo. 2018. ``Quanteda: An R Package for
the Quantitative Analysis of Textual Data.'' \emph{Journal of Open
Source Software} 3 (30): 774. \url{https://doi.org/10.21105/joss.00774}.

\leavevmode\hypertarget{ref-Bodnaruk2013}{}%
Bodnaruk, Andriy, Tim Loughran, and Bill McDonald. 2013. ``Using 10-K
Text to Gauge Financial Constraints.'' \emph{Ssrn} 50 (4): 623--46.
\url{https://doi.org/10.2139/ssrn.2331544}.

\leavevmode\hypertarget{ref-Bourveau2018}{}%
Bourveau, Thomas, Yun Lou, and Rencheng Wang. 2018. ``Shareholder
Litigation and Corporate Disclosure: Evidence from Derivative
Lawsuits.'' \emph{Journal of Accounting Research} 56 (3): 797--842.
\url{https://doi.org/10.1111/1475-679X.12191}.

\leavevmode\hypertarget{ref-Enev2017}{}%
Enev, Maria. 2017. ``Going Concern Opinions and Management's Forward
Looking Disclosures: Evidence from the MD\&A.''
\url{https://doi.org/10.2139/ssrn.2938703}.

\leavevmode\hypertarget{ref-Feldman2008}{}%
Feldman, Ronen, Suresh Govindaraj, Joshua Livnat, and Benjamin Segal.
2008. ``The Incremental Information Content of Tone Change in Management
Discussion and Analysis.'' \url{https://doi.org/10.2139/ssrn.1126962}.

\leavevmode\hypertarget{ref-Feldman2010}{}%
---------. 2010. ``Management's tone change, post earnings announcement
drift and accruals.'' \emph{Review of Accounting Studies} 15 (4):
915--53. \url{https://doi.org/10.1007/s11142-009-9111-x}.

\leavevmode\hypertarget{ref-fisher2016natural}{}%
Fisher, Ingrid E, Margaret R Garnsey, and Mark E Hughes. 2016. ``Natural
Language Processing in Accounting, Auditing and Finance: A Synthesis of
the Literature with a Roadmap for Future Research.'' \emph{Intelligent
Systems in Accounting, Finance and Management} 23 (3): 157--214.

\leavevmode\hypertarget{ref-gandhi2019using}{}%
Gandhi, Priyank, Tim Loughran, and Bill McDonald. 2019. ``Using Annual
Report Sentiment as a Proxy for Financial Distress in Us Banks.''
\emph{Journal of Behavioral Finance} 20 (4): 424--36.

\leavevmode\hypertarget{ref-Gandhi2017}{}%
---------. 2017. ``Using Annual Report Sentiment as a Proxy for
Financial Distress in U.S. Banks.'' \emph{Ssrn}, March, 1--13.
\url{https://doi.org/10.2139/ssrn.2905225}.

\leavevmode\hypertarget{ref-Hajek2015}{}%
Hájek, Petr, and Vladimír Olej. 2015. ``Word categorization of corporate
annual reports for bankruptcy prediction by machine learning methods.''
In \emph{Lecture Notes in Computer Science (Including Subseries Lecture
Notes in Artificial Intelligence and Lecture Notes in Bioinformatics)},
9302:122--30. \url{https://doi.org/10.1007/978-3-319-24033-6_14}.

\leavevmode\hypertarget{ref-hillegeist2004assessing}{}%
Hillegeist, Stephen A, Elizabeth K Keating, Donald P Cram, and Kyle G
Lundstedt. 2004. ``Assessing the Probability of Bankruptcy.''
\emph{Review of Accounting Studies} 9 (1): 5--34.

\leavevmode\hypertarget{ref-huizinga2012bank}{}%
Huizinga, Harry, and Luc Laeven. 2012. ``Bank Valuation and Accounting
Discretion During a Financial Crisis.'' \emph{Journal of Financial
Economics} 106 (3): 614--34.

\leavevmode\hypertarget{ref-Lopatta2017}{}%
Lopatta, Kerstin, Mario Albert Gloger, and Reemda Jaeschke. 2017. ``Can
Language Predict Bankruptcy? The Explanatory Power of Tone in 10-K
Filings.'' \emph{Accounting Perspectives} 16 (4): 315--43.
\url{https://doi.org/10.1111/1911-3838.12150}.

\leavevmode\hypertarget{ref-LopezLira2019}{}%
Lopez Lira, Alejandro. 2019. ``Risk Factors That Matter: Textual
Analysis of Risk Disclosures for the Cross-Section of Returns.''
\url{https://doi.org/10.2139/ssrn.3313663}.

\leavevmode\hypertarget{ref-lopucki2006bankruptcy}{}%
LoPucki, Lynn M. 2006. ``Bankruptcy Research Database.''

\leavevmode\hypertarget{ref-Loughran2009}{}%
Loughran, Tim, and Bill Mcdonald. 2009. ``Plain English , Readability ,
and 10-K Filings.'' \emph{English}.

\leavevmode\hypertarget{ref-Loughran2011}{}%
Loughran, T I M, and Bill Mcdonald. 2011. ``When is a Liability not a
Liability ? Textual Analysis , Dictionaries , and 10-Ks Journal of
Finance , forthcoming.'' 1. Vol. 66.
\url{https://doi.org/10.1111/j.1540-6261.2010.01625.x}.

\leavevmode\hypertarget{ref-Mayew2015}{}%
Mayew, William J., Mani Sethuraman, and Mohan Venkatachalam. 2015.
``MD\&A disclosure and the firm's ability to continue as a going
concern.'' \emph{Accounting Review} 90 (4): 1621--51.
\url{https://doi.org/10.2308/accr-50983}.

\leavevmode\hypertarget{ref-nguyen2020textual}{}%
Nguyen, Ba-Hung, and Van-Nam Huynh. 2020. ``Textual Analysis and
Corporate Bankruptcy: A Financial Dictionary-Based Sentiment Approach.''
\emph{Journal of the Operational Research Society}, 1--20.

\leavevmode\hypertarget{ref-pennebaker2015development}{}%
Pennebaker, James W, Ryan L Boyd, Kayla Jordan, and Kate Blackburn.
2015. ``The Development and Psychometric Properties of Liwc2015.''

\leavevmode\hypertarget{ref-pennebaker2001linguistic}{}%
Pennebaker, James W, Martha E Francis, and Roger J Booth. 2001.
``Linguistic Inquiry and Word Count: LIWC 2001.'' \emph{Mahway: Lawrence
Erlbaum Associates} 71 (2001): 2001.

\leavevmode\hypertarget{ref-rajan2015failure}{}%
Rajan, Uday, Amit Seru, and Vikrant Vig. 2015. ``The Failure of Models
That Predict Failure: Distance, Incentives, and Defaults.''
\emph{Journal of Financial Economics} 115 (2): 237--60.

\leavevmode\hypertarget{ref-Shirata2011}{}%
Shirata, Cindy Yoshiko, Hironori Takeuchi, Shiho Ogino, and Hideo
Watanabe. 2011. ``Extracting Key Phrases as Predictors of Corporate
Bankruptcy: Empirical Analysis of Annual Reports by Text Mining.''
\emph{Journal of Emerging Technologies in Accounting} 8 (1): 31--44.
\url{https://doi.org/10.2308/jeta-10182}.

\leavevmode\hypertarget{ref-sobehart2001measuring}{}%
Sobehart, Jorge, and Sean Keenan. 2001. ``Measuring Default
Accurately.'' \emph{Risk} 14 (3): 31--33.

\leavevmode\hypertarget{ref-Tao2018}{}%
Tao, Jie, Amit V. Deokar, and Ashutosh Deshmukh. 2018. ``Analysing
forward-looking statements in initial public offering prospectuses: a
text analytics approach.'' \emph{Journal of Business Analytics} 1 (1):
54--70. \url{https://doi.org/10.1080/2573234x.2018.1507604}.

\leavevmode\hypertarget{ref-tinoco2013financial}{}%
Tinoco, Mario Hernandez, and Nick Wilson. 2013. ``Financial Distress and
Bankruptcy Prediction Among Listed Companies Using Accounting, Market
and Macroeconomic Variables.'' \emph{International Review of Financial
Analysis} 30: 394--419.

\leavevmode\hypertarget{ref-wu2010comparison}{}%
Wu, Yanhui, Clive Gaunt, and Stephen Gray. 2010. ``A Comparison of
Alternative Bankruptcy Prediction Models.'' \emph{Journal of
Contemporary Accounting \& Economics} 6 (1): 34--45.

\leavevmode\hypertarget{ref-Yang2018}{}%
Yang, Fang, Burak Dolar, and Lun Mo. 2018. ``Textual Analysis of
Corporate Annual Disclosures: A Comparison between Bankrupt and
Non-Bankrupt Companies.'' \emph{Journal of Emerging Technologies in
Accounting} 15 (1): 45--55. \url{https://doi.org/10.2308/jeta-52085}.
\end{cslreferences}

\bibliographystyle{unsrt}
\bibliography{edgar.bib}

\end{document}